\documentclass[conference]{IEEEtran}

\usepackage{cite}
\usepackage{amsmath,amssymb,amsfonts}
\usepackage{booktabs}
\usepackage{multirow}
\usepackage{graphicx}
\usepackage{xcolor}
\usepackage{stfloats}
\usepackage{subcaption}
\usepackage{algorithm}
\usepackage{algpseudocode}
\usepackage{diagbox}
\usepackage{seqsplit}
\usepackage{soul}
\usepackage{balance}
\usepackage{url}

\definecolor{red}{HTML}{de2d26}
\definecolor{green}{HTML}{31a354}

\AtBeginDocument{%
  \providecommand\BibTeX{{%
    Bib\TeX}}}

\def\BibTeX{{\rm B\kern-.05em{\sc i\kern-.025em b}\kern-.08em
    T\kern-.1667em\lower.7ex\hbox{E}\kern-.125emX}}

\begin{document}

\title{The Impact of Train-Test Leakage on Machine Learning-based Android Malware Detection}


\author{\IEEEauthorblockN{Guojun Liu}
  \IEEEauthorblockA{
    University of South Florida\\
    Tampa, Florida\\
    guojunl@usf.edu}
\and
\IEEEauthorblockN{Doina Caragea}
\IEEEauthorblockA{
  Kansas State University\\
  Manhattan, Kansas\\
  dcaragea@ksu.edu}
\and
\IEEEauthorblockN{Xinming Ou}
  \IEEEauthorblockA{
    University of South Florida\\
    Tampa, Florida\\
    xou@usf.edu}
\and
\IEEEauthorblockN{Sankardas Roy}
\IEEEauthorblockA{
  Bowling Green State University\\
  Bowling Green, Ohio\\
  sanroy@bgsu.edu}
}

\maketitle

\begin{abstract}
  When machine learning is used for Android malware detection, an app
needs to be represented in a numerical format for training and testing.
We identify a widespread occurrence of
distinct Android apps that have identical or nearly identical app
representations. In particular, among app samples in the testing
dataset, there can be a significant percentage of apps that have an
identical or nearly identical representation to an app in the training
dataset. This will lead to a data leakage problem that inflates a
machine learning model's performance as measured on the testing
dataset. The data leakage not only could lead to overly optimistic
perceptions on the machine learning models' ability to generalize
beyond the data on which they are trained, in some cases it could also
lead to qualitatively different conclusions being drawn from
the research. We present two case studies
to illustrate this impact. In the first case study, the data leakage
inflated the performance results but did not impact the overall
conclusions made by the researchers in a qualitative way. In the
second case study, the data leakage problem would have led to
qualitatively different conclusions being drawn from the research.
We further examine the real-world impact of the data leakage by
dissecting the capability of memorization and the capability of generalization
of a machine learning model, and show that by removing leakage from
testing data, the evaluation results better reflect the machine
learning model's utility in real-world Android malware detection scenarios.
  
\end{abstract}


\section{Introduction}
\label{sec:intro}

\IEEEPARstart{M}{achine} learning (ML) has been extensively researched for its use in
Android malware detection.  In such research a raw app is transformed
into a suitable numeric representation -- a feature vector, a
multi-dimensional latent representation, a graph structure, etc., each
tailored to a specific ML model. The transformation is typically a
lossy function and distinct apps may result in identical app
representations.  This could lead to a substantial number of app
samples in the testing dataset that have representations identical or
nearly identical to representations of some samples in the training
dataset, a phenomenon known as a type of data
leakage~\cite{Kapoor:Patterns23, Elangovan:arXiv2021, Tampu:SD22,
Florensa:NARGAB24}.

Data leakage has been a recognized phenomenon in ML applications for a
long time~\cite{Nisbet:Handbook2009}. Early research defined different
types of data leakage~\cite{Kaufman:TKDD12, Ghani:top10leakage,
Soni:Medium2019} which may arise at various stages of ML model
development and application.  Data leakage has been identified as a
significant concern in the implementation of ML algorithms across many
scientific fields~\cite{Kapoor:Patterns23, Zhu:EnvironmentScience23,
Elangovan:arXiv2021, Tampu:SD22, Florensa:NARGAB24}, including in
computer security~\cite{Arp:USENIXSeurity22}.  We focus on the leakage
caused by identical or nearly identical app representations between
training and testing datasets, a type of leakage that has been largely
unexplored in the security research community. In machine learning, a
model learned on a training set shall be tested on data that is
disjoint from the training data. If an app in the testing set has a
representation that is identical to that of an app in the training,
the model is tested on data it has already seen during training, and
this is called train-test leakage in machine learning literature.
Note that such leakage could arise from the inherent nature of the
data rather than errors in experimental design.  The leakage generally
results in inflated performance results that do not reflect the
model's capacity to generalize, since it could simply memorize the
leaked data to make a prediction. Some identify the problem as
conflating the model's ability to memorize with its ability to
generalize~\cite{Elangovan:arXiv2021}. Even if the representation is
not completely identical, but nearly identical, it still calls into
question whether the performance measured on the testing data truly
reflects the model's generalizability beyond the data it is trained
on.  Our framing of the train-test leakage problem follows that used
in machine learning literature of domains such as
bioinformatics~\cite{Florensa:NARGAB24}, environmental
research~\cite{Zhu:EnvironmentScience23}, medical
imaging~\cite{Tampu:SD22}, and NLP~\cite{Elangovan:arXiv2021}.

Past research on Android malware detection has adopted various
approaches to constructing train and test data for machine learning
models.  Recent work by Pendlebury et al.~\cite{Pendlebury:USENIX19}
pointed out biases arising from train/test data construction and how
to avoid them.  Our analysis found that even when one follows those
best practices in constructing training and testing datasets, there
could still be large fractions of apps in testing data that share the
same feature vectors as apps in training data, leading to data
leakage.  We ask the questions: 1) how will such leakage impact an ML
model's measured performance, and 2) given that such leakage will
happen ``naturally'' when the model is applied to real-world data, is
there any practical value in discerning such differences of
performance measurement?

To answer these questions, we present two case studies of recent
research efforts of using ML in Android malware detection. The first
is that of \mbox{Chen et al.~\cite{Chen:USENIX23}}, which investigated
the combination of contrastive and active learning in Android malware
detection. The second is from the authors of this paper, where we
investigate the application of graph neural networks to Android
malware detection.  Both works divide training and testing data
temporally so that training data pre-date testing data. In both cases,
we found that substantial fractions of apps in the testing data have
identical or nearly identical representations to apps in the
corresponding training data. These apps in the testing data are
considered leakage based on two different criteria, depending on the
specific ML model used.

\begin{enumerate}
\item For machine learning models that train on feature vectors
extracted from apps, we consider a testing app as leakage if its
feature vector is identical to the feature vector of at least one app
in the corresponding training data.

\item For machine learning models that analyze app structure through
graph representations, these graphs are first converted into numerical
latent representations called ``graph embeddings'' for the final
detection stage. To identify data leakage in these models, we leverage
cosine similarity, a well-established metric for comparing embeddings.
If a testing app's embedding has a high cosine similarity score with
at least one training app's embedding, we consider the testing app as
leakage.
\end{enumerate}

For both Chen et al.~\cite{Chen:USENIX23}'s work and our own work, we
evaluate the machine learning models' performance in two ways: once
using the original testing data in the research, and again using the
same data but after removing apps identified as data leakage. Our
evaluation reveals that removing leakage from the testing data impacts
all models from both works. In Chen et al.~\cite{Chen:USENIX23}'s
work, all evaluated models experience an average decrease in
performance. Similarly, in our work, seven out of eight experiments
show a performance decline.  In both works, the researchers' main
objectives are to compare various models' performance. While the
models' average performances dropped when the leakage data was removed
from the testing data, we investigated if the comparative results
between the models may also be impacted. We found that in Chen et
al.~\cite{Chen:USENIX23}'s work, the overall conclusions about the
comparative advantage of the proposed contrastive learning+active
learning approach still hold, although the extent of improvement did
shrink considerably in one of the two datasets. Thus, in this case,
the train-test leakage did not impact the overall conclusion from the
research in a qualitative way.  For our own work on comparing graph
neural networks' capability with a traditional random forest model,
the graph neural network model's performance is inferior to that of
the random forest model when measured on the original testing data.
After the leakage data is removed from the testing data, the graph
neural network's performance exceeds that of random forest. In this
case, the train-test leakage would have led to a qualitatively
different conclusion regarding the comparative advantage of the graph
neural network model against the random forest model.

In practice a detector needs to make a prediction for every app,
regardless of whether the app has been seen before or not. Given this
fact, does discerning performance measurement differences after
removing the leakage app provide any beneficial insights for operational decision-making?
This is especially important in the second case study --
does the new measurement suggest favoring graph neural networks over random forests for operational use?
How does this reconcile with the fact that random
forests performs better than graph neural network on the complete
testing data which comprises all apps necessitating processing?

To answer this question, we observe that in practice, machine learning
is just one tool out of many analysts rely on to make a determination
on an app's maliciousness. For apps that are known malware, i.e., already
labeled as malware in the training data, there is likely other more
efficient approach to detect them, e.g., by using a known signature.
Only when a malware is unknown, i.e., corresponding to the non-leakage
portion of the testing data, a machine learning model is potentially
more useful in providing guidance. Thus testing a machine learning
model's performance on the non-leakage portion of the testing data
tells more about the model's utility in practice.

In summary, our research investigates the impact of train-test leakage
on the evaluation of machine learning-based Android malware detection
systems.  In general the measured performance drops when the leakage
is removed. Identifying and removing such leakage is not only
important in showing machine learning's true capability of
generalizing to substantially different Android apps that it has not
been trained on, but also has practical implication on how to apply
machine learning in this domain.

\section{Background} 
\label{background}

Both traditional machine learning (ML) and deep learning (DL)
frameworks have been proposed and extensively investigated for Android
malware detection and demonstrated remarkable performance over the
past decade. We provide some background relevant to the discussion in
this paper.

\subsection{App Representation for Machine Learning}
\label{sec:feature-vector}

Vector representation is a common form of input to an ML model.
Creating a feature/term vector typically involves extracting specific
features from an app, which can be static features, dynamic features,
or both. Once the features have been extracted, an app can be
transformed into a binary vector, term frequency (TF) vector or a term
frequency-inverse document frequency (TF-IDF) vector. Among these, the
binary vector is the most widely used in ML-based Android malware
detection. Here each vector component corresponds to a feature with
$1$ denoting the presence of the feature and $0$ denoting its
absence.\footnote{While one could design feature vectors to use
general TF counts/integers or TF-IDF floating numbers, we have not
seen much research that adopts these feature representations for
Android malware detection. The binary $\{0, 1\}$ vectors as described
here are used almost universally as representations in the Android
malware detection literature.}

DL-based Android malware detection has also used vectorized app
representations as model inputs~\cite{Yuan:SIGCOMM14, Su:TrustCom16,
McLaughlin:CODASPY17, Karbab:DI18}.  However, DL's input capability
extends beyond one-dimensional, allowing for richer multi-dimensional tensor
representations of text, image, graph, and other structured data as
inputs. Furthermore, DL has the ability to automatically learn latent
vector representations (called embeddings) that capture high-level
semantic relationships in its inputs~\cite{Goodfellow:DL2016,
Kipf:LCLR17}.  Prior works have transformed raw bytecode of an APK's
DEX files into a grayscale or color image suitable for convolutional
neural networks, which subsequently learn feature representations at
many levels~\cite{Huang:BigData18, Daoudi:DMLSD21}.  Other works have
represented an app as method-level Control Flow Graphs (CFGs) or Data
Flow Graphs (DFGs), which can then be fed into convolutional neural
networks or graph convolutional network
architectures~\cite{Xu:ICFEM18, John:ISEA20, Gao:CS21}.

\subsection{Experimental Biases}
\label{sec:biases}

Despite significant advances in machine learning based Android malware
detection, experimental biases and pitfalls persist, demanding careful
consideration. \mbox{Pendlebury et al.~\cite{Pendlebury:USENIX19}}
highlighted two crucial concerns: 1) spatial bias: test data does not
reflect real-world ratio of benign to malicious apps, and 2) temporal
bias: apps in training and testing do not follow a realistic timeline.
One example of temporal bias is when some apps in the training set are
more recent than those in the testing set, essentially using future
apps to predict the behavior of earlier apps. This bias often
occurs when a large amount of data is collected from single or
multiple sources and randomly split into training and testing sets
without considering the release date of the apps.  Models evaluated on
such temporally biased data cannot reflect its utility in operation,
where such future knowledge is impossible to obtain.  Furthermore,
Arp~et~al.~\cite{Arp:USENIXSeurity22} identified sampling bias as
another pitfall in learning-based security systems, including Android
malware detection. This bias occurs when sampling inadvertently
introduces artifacts that do not exist in real-world data.  For
example, if benign apps are primarily collected from one market, while
malicious apps are mainly collected from others, the model may end up
learning the difference between the two markets instead of the
difference between malware and benign apps. Dataset collection needs
to be carefully done to avoid these biases.

\section{Train-Test Leakage in Android Malware Detection}
\label{sec:leakage}

For machine learning models, if the representation of an Android app
in the testing set coincides with the representation of another app in
the training set, the app in the testing set is considered leakage --
the model will be tested on an app that the model was also trained on.
Even though the underlying apps are distinct, since an ML model only
learns from the app's representation, identical representations are
the same data to the ML model.  Train-test leakage has been
extensively discussed for machine learning models applied in other
domains, including bioinformatics~\cite{Florensa:NARGAB24},
environmental research~\cite{Zhu:EnvironmentScience23}, medical
imaging~\cite{Tampu:SD22}, and NLP~\cite{Elangovan:arXiv2021}.  It is
well known that such data leakage will inflate performance results, especially if the fraction of leakage is significant.  The main
problem is that performance metrics measured on the testing set
containing leaked data conflates the model's ability to generalize
with its ability to memorize. For the portion of apps in the testing data that
are identical to some in training data, the model can simply memorize
those data's label as ``prediction'' result.  This could undermine the
validity of conclusions drawn from ML-based research, especially if
the conclusion is sensitive to a model's ability to generalize to data
it has not seen in training. In these domains, researchers have
addressed the train-test leakage by re-dividing the train/test splits
to avoid such leakage. This is not an option for most security
problems, and for Android malware detection in particular, since doing
so will likely introduce temporal biases as discussed in
Section~\ref{sec:biases}.  To assess the impact of data leakage on
evaluation results, we can exclude the ``leaked apps'' from the
testing dataset, re-evaluate the models on the ``cleaned'' testing
data, re-draw all conclusions thereafter, and compare with the results
before the leakage removal.  To the best of our knowledge, there has
not been any study that directly examines train-test leakage's impact
on evaluating machine learning models' utility in a security problem
domain. One recent manuscript~\cite{PrimeVul:arXiv2024} identifies
train-test leakage in vulnerability datasets for evaluating code
language models, but did not isolate the leakage's impact on
evaluating the models' performance.

Another type of data leakage, which is sometimes confused with
train-test leakage, is called temporal
leakage~\cite{Kapoor:Patterns23}. Temporal leakage arises from
temporal bias as discussed in Section~\ref{sec:biases}. It happens
when testing data pre-dates training data, essentially requiring a
model to be trained on future data not available at the time of model
usage. Temporal leakage is always a flaw in experiment design, since
it is evaluating a model under a scenario that can never materialize
in real world. Train-test leakage, on the other hand, could very well
happen in a realistic model application scenario. For example, due to
the nature of Android app evolvement, apps released today may bear
similarity to some apps released in the past, creating possible
train-test leakage when those older apps are used for training.  The
problem caused by train-test leakage is the conflation of model's
ability to generalize with its ability to memorize, leading to
potentially inaccurate conclusion drawn about a machine learning
model's utility in practice. {\bf This paper's sole focus is
train-test leakage. The two case studies we use are both free of
temporal leakage.}

\subsection{Evaluation Metrics}
\label{sec:metrics}

F1 score is a common metric adopted by Android malware detetion work,
but it is sensitive to the malware/benign app
ratio~\cite{Pendlebury:USENIX19}.  When we remove leakage apps from
testing set to examine the leakage apps' impact on the model's
performance measurement, we may also have changed the malware ratio in
the testing set, which may impact F1 score by itself.  For isolating
the impact of train-test leakage on the performance measurement, we
use balanced accuracy (BA), defined as the arithmetic mean of
sensitivity and specificity. The full definitions of the various
evaluation metrics can be found in Appendix A.  Unlike F1 score, BA is
insensitive to class ratio, which will allow us to show the impact of
the leakage on the performance measurement.  We can compare an ML
model's BA on the testing data before and after the leakage is
removed.  A performance decrease after removing the leakage indicates
that the leaked apps in the test data inflated the performance results
of the model.

In the next two sections, we present two case studies of research
where ML is used for Android Malware detection. We examine the impact
of leaked apps in testing data on the performance results, and discuss
how this may affect the ultimate conclusions the research aims to
draw. When comparing the performance results before and after leakage
removal, we include both balanced accuracy (BA) and the original
metrics adopted by the researchers (e.g., F1 score).

\section{Case Study 1: Continuous Learning for Android Malware Detection}
\label{sec:case_chen}

  Chen et al.'s work~\cite{Chen:USENIX23} is one of the most recent
studies that apply machine learning to Android malware
detection. The authors made all research artifacts publicly available,
including the datasets used in the experiments and the
model-related code. We were able to reproduce the results published in
the paper using the published code and data without any trouble. This
enabled us to examine the effect of train-test data leakage on this
work.

\subsection{Summary of the Research and Datasets Used}
\label{sec:summary_chen}

\begin{figure}[h]
     \centering
     \includegraphics[width=0.5\textwidth]{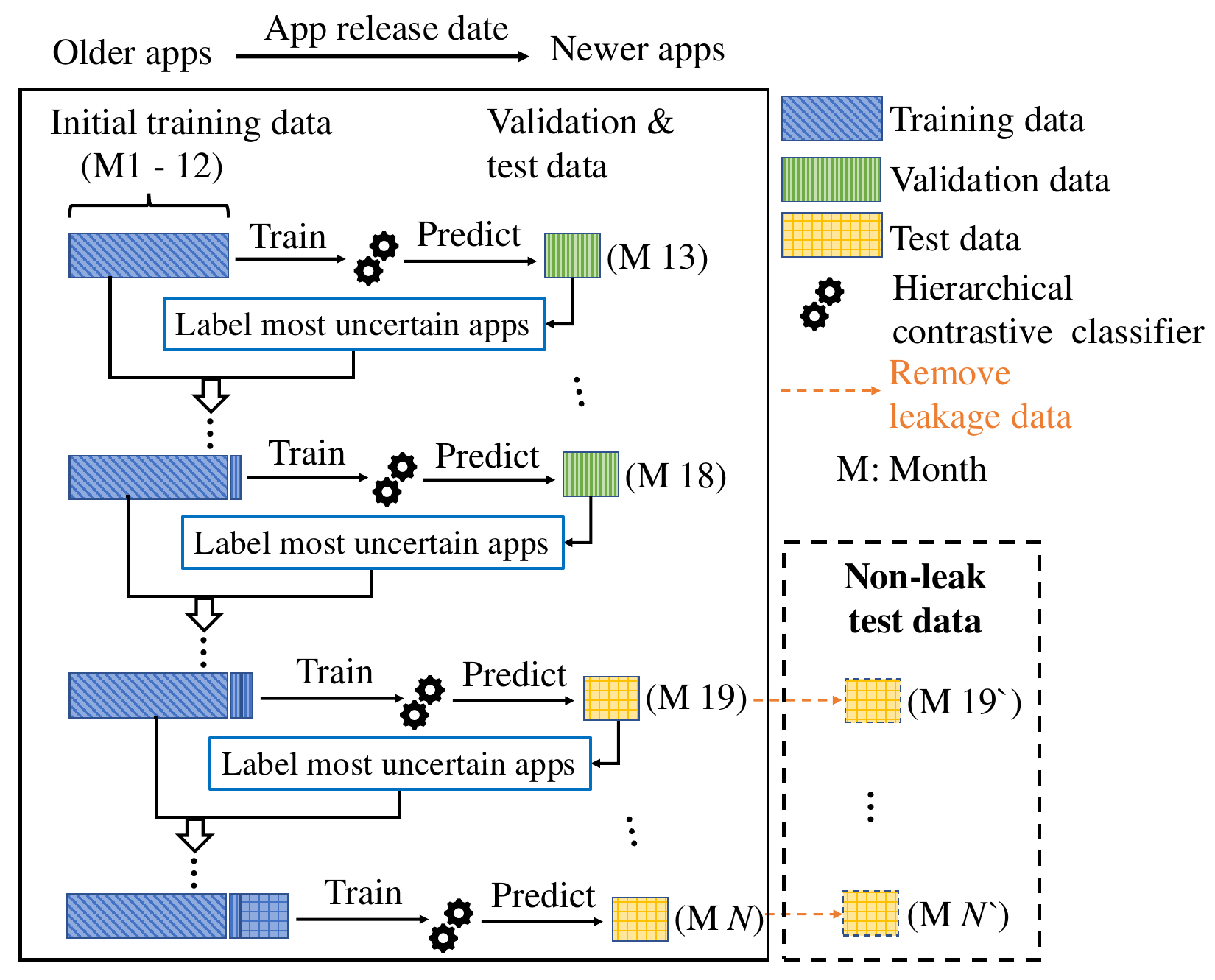}
     \caption{The left rectangle illustrates Chen et
al.'s~\cite{Chen:USENIX23} continuous learning approach.  Each month
new samples are added to the training data through active learning; a
new hierarchical contrastive classifier is trained using warm start
and evaluated on next month's data. Month 13-18 are used as validation
data. Test data starts from month 19. The dotted rectangle on the
right illustrates removing leakage apps from test data for
evalaution. The non-leak evaluation does not change Chen et al.'s
original approach; it simply removes leakage apps from testing data
for evaluation.  }
     \label{fig:continuous_learning}
 \end{figure}

Chen et al.~\cite{Chen:USENIX23} proposed a continuous learning
approach that combines contrastive learning and active learning to
improve ML-based Android malware detector's capability of adapting to
concept drift~\cite{Jordaney:USENIX17}.  To address the challenge
posed by evolving malware techniques, the researchers introduced a
hierarchical contrastive learning scheme and a novel sample selection
technique for continuous learning as shown in
Figure~\ref{fig:continuous_learning}. The process begins by using the
apps from the first year of the dataset as the initial training data
to build the initial hierarchical contrastive classifier. Then an
active learning process starts. The trained classifier is used to
predict the labels of next month's samples. The classifier computes an
uncertainty score for all samples in the month (based on contrastive
learning embeddings) and selects the apps with the most uncertain
scores, based on a labeling budget. These selected apps are then
labeled and added into the training set for the next iteration, where
a new model is trained using warm start (i.e., starting with the model
from the prior iteration).  This process repeats and the model is
continuously trained and tested on each next month's apps. The
ultimate goal is to minimize human labeling effort, while keeping the
model updated with evolving malware.

The hierarchical contrastive classifier proposed by the authors
consists of two subnetworks. The first subnetwork is an encoder (Enc),
which encodes the original inputs using contrastive learning based on
malware family information. The second subnetwork is a multilayer
perceptron (MLP), which uses the encoded input to predict a binary
label. This model architecture is denoted as \mbox{Enc+MLP}. The
authors compared their approach with several active learning baselines
using both binary and multi-class classifiers. The binary classifiers
include a fully connected neural network~(NN), a linear support vector
machine (SVM), and gradient boosted decision trees (GBDT), and the
multi-class classifiers (which predict malware families)\footnote{The
authors note that the output of the multi-class classifiers is
converted to binary classes (malware/benign) to allow for a fair
evaluation of all approaches used.} include multiplayer perceptron
(MLP) and SVM. An additional experiment investigated a ``Multi-class
MLP + Binary SVM'' architecture.

Chen et al.~\cite{Chen:USENIX23} compared their continuously learning
approach against these prior classifiers combined with several active
learning schemes.  Two datasets were used in the evaluation:
APIGraph~\cite{Zhang:CCS20}, with app years 2012-2018, and
AndroZoo~\cite{Allix:MSR16}, with app years 2019-2021. The apps are
represented as feature vectors, using DREBIN~\cite{Arp:NDSS14}
features extracted from the apps, which capture an app's access to
hardware, requested permissions, component names, intents, restricted
API calls, used permissions, suspicious API calls, and network
addresses.  The number of DREBIN features used in the APIGraph dataset
is 1,159, while the number of features used in the AndroZoo dataset is
16,978~\cite{Chen:USENIX23}.  For the APIGraph dataset, the initial
model was trained on year 2012 data, and validation was performed on
data covering the months from 2013-01 to 2013-06.  The testing data
included the months from 2013-07 to 2018-12. For the AndroZoo dataset,
the initial model was trained on year 2019 data, while validation was
performed on the months from 2020-01 to 2020-06, and testing included
the months from 2020-07 to 2021-12.  This arrangement of
training/validation/testing data ensured that no temporal leakage was
introduced.  The authors used the metrics of FNR, FPR, and F1
score. In both datasets, the performance metrics were averaged across
all testing data months.  The results show that their approach
outperforms all prior works.

 \begin{figure*}[h]
     \centering
     \begin{subfigure}{0.49\textwidth}
         \centering
         \includegraphics[width=\textwidth]{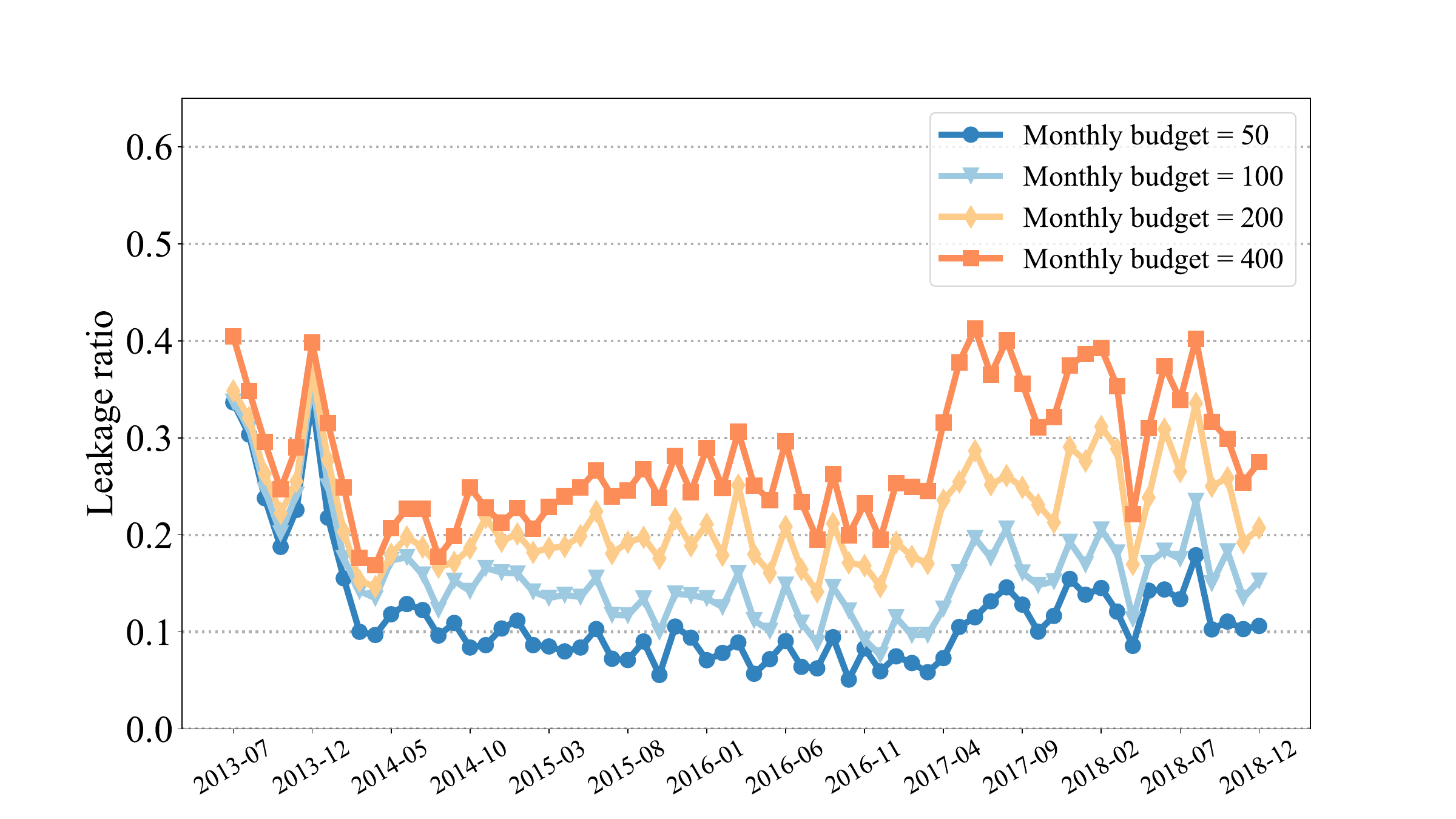}
         \caption{APIGraph test data (2013-07 to 2018-12)}
         \label{fig:apigraph_identical_vec}
     \end{subfigure}
     \hfill
     \begin{subfigure}{0.49\textwidth}
         \centering
         \includegraphics[width=\textwidth]{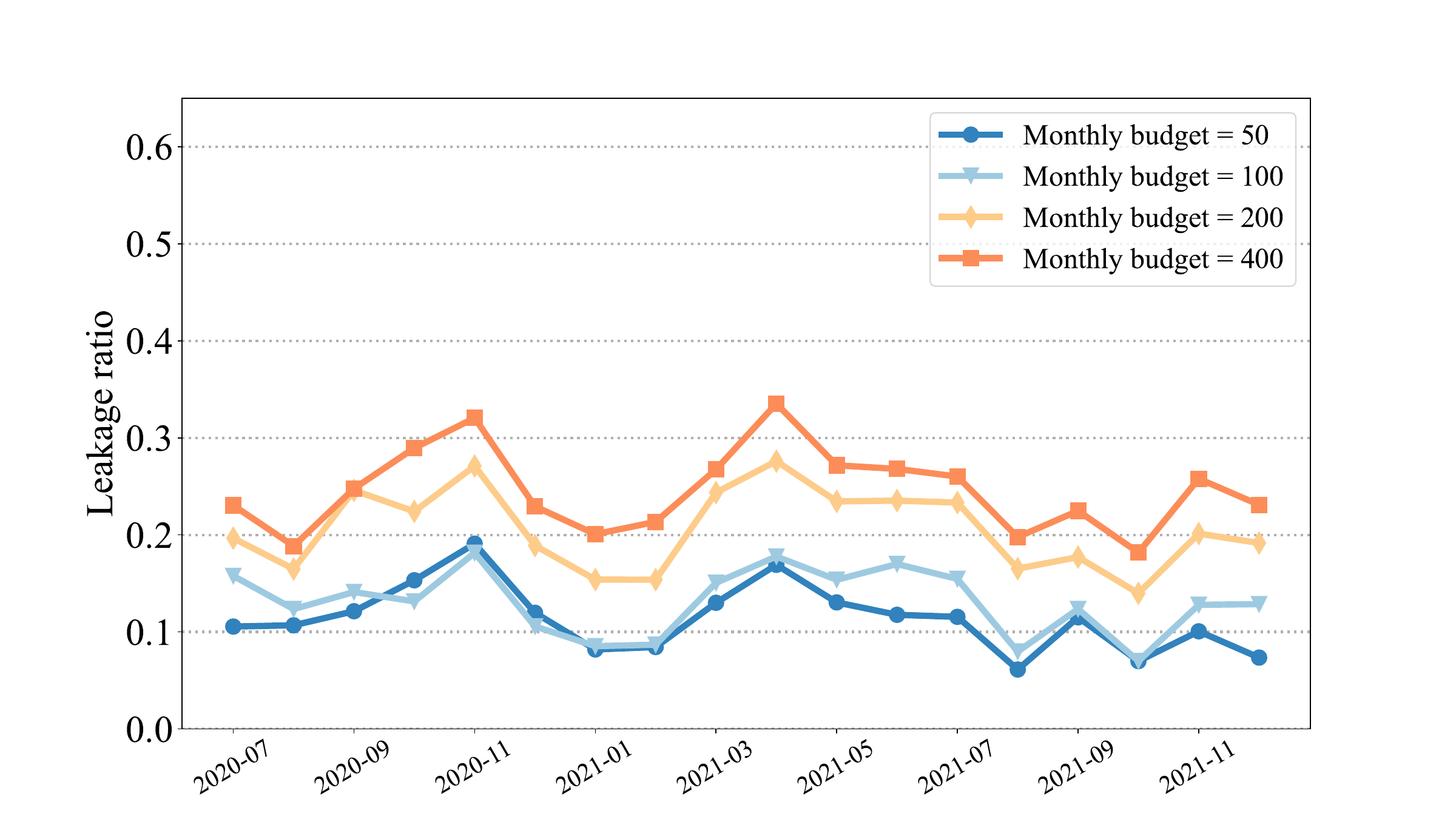}
          \caption{AndroZoo test data (2020-07 to 2021-12)}
         \label{fig:androzoo_identical_vec}
       \end{subfigure}
       \caption{Leakage ratio in APIGraph and AndroZoo datasets used in Chen et al.'s work}
      \label{fig:identical_vec_ratio}
  \end{figure*}

\subsection{Identification of Train-Test Leakage}
\label{sec:leak_chen}

In identifying train-test leakage in the dataset, we consider a test
app a leakage if its feature vector is identical to that of at least
one app in the corresponding training data. In the continuous learning
approach, new samples are added to the training data each month, and
the next month's data is used for testing. Thus, to identify leakage
in a testing month, we first need to track all the newly labeled
samples added to the training data in the previous months. We then
compare the apps in the testing data against all the apps in the
initial training data (first 12 months) plus the newly labeled apps
that have been added so far. Any app in the testing month that share
the same feature vector with at least one of these training apps is
identified as leakage.

Chen et al.~\cite{Chen:USENIX23} evaluated seven different models,
combined with several active learning schemes for selecting uncertain
apps for labeling and adding into training data.  Thus, for the same
testing month, the training data used depends on which model is being
evaluated and what sample selection scheme is used in active
learning. All models use the first 12 months of data as initial
training, but the newly added samples for subsequent training will
differ across the different models. In effect, for the same testing
month, each model being evaluated gives rise to a distinct subset that
is considered leakage. To examine the impact on a model's performance,
we would need to remove the model-dependent leakage apps.  This poses
a problem for comparing the performance across the models: the testing
data is no longer the same for those models after the leakage is
removed, since what is removed may differ from one model to another.
This will make it impossible to compare the performance results of
those models, since such comparison ought to be done on the same
testing data.

To allow us to compare the various models after removing leakage, we
must adhere to two principles:

\begin{enumerate}
\item All leakage apps in the testing month must be removed for all models.
  
\item All models must be evaluated on the same testing data.
\end{enumerate}

It then follows that we must remove from the testing data the
\textit{union} of the leakage apps for all models, to allow us to
compare the results among those models after leakage removal.  The
dotted rectangle at the right corner of
Figure~\ref{fig:continuous_learning} illustrates the removal of the
union of the leakage apps for all models from a testing month. We
calculate the ratio of the removed apps over the total number of apps
in that test month.  Figure~\ref{fig:apigraph_identical_vec} and
Figure~\ref{fig:androzoo_identical_vec} illustrate this leakage ratio
for each testing month.  Higher monthly labeling budget leads to more
apps added to the training data over time, and hence higher chance a
testing app's feature vector being identical to one of the training
apps. Overall about 10\%-40\% of testing apps are found to be leakage
for the APIGraph test data, and about 10\%-30\% of testing apps are
found to be leakage for the AndroZoo test data.

It is important to note that the removal of leakage apps is only done
for the purpose of evaluating the trained models. We do not change how
the active learning scheme works in Chen et al.'s work. For each
month, the total set of apps in that month is used for identifying the
most uncertain predictions, just as what the authors did in their
experiment. We simply provide another measurement of the models'
performance -- running the same models on the testing data with
leakage removed, and compare this performance against the performance
measured on the unchanged testing data.

\subsection{Impact of the Leakage on Performance Results}
\label{sec:experiment_chen}

\begin{table*}[h]
  \centering
  \caption{Results Before and After Removing Leak from Test Data in Chen et al.'s Work -- APIGraph Dataset}
  \label{tab:eval_diff_apigraph}
  \begin{tabular}{c|c|cccc||cccc}
    \toprule[1.6pt]
    \multirow{3}{*}{\begin{tabular}[c]{@{}c@{}}\textbf{Monthly}\\ \textbf{Sample}\\ \textbf{Budget}\end{tabular}} & \multirow{3}{*}{\begin{tabular}[c]{@{}c@{}}\textbf{Model}\\ \textbf{Architecture}\end{tabular}}  & \multicolumn{4}{c||}{\multirow{2}{*}{\begin{tabular}[c]{@{}c@{}}\textbf{Measurement on Original Data}\\ Average Performance (\%)\end{tabular}}} & \multicolumn{4}{c}{\multirow{2}{*}{\begin{tabular}[c]{@{}c@{}}\textbf{Measurement after Leak Removed from Test}\\ Average Performance (\%)\end{tabular}}}\\                                               & &  \multicolumn{4}{c||}{} & \multicolumn{4}{c}{} \\
    & &  FNR & FPR & F1 & BA & FNR & FPR & F1 & BA \\
    \hline\hline
    \multirow{8}{*}{50}
    & Binary MLP &  23.99 & 0.56 & 83.37 & 87.72 & 27.44 & 0.54 & 80.47 & 86.01\\
    \cline{2-10}
    & Multiclass MLP &  14.57 & 5.58 & 72.57 & 89.92 & 18.27 & 4.33 & 69.38 & 88.70\\
    \cline{2-10}
    & \begin{tabular}[c]{@{}c@{}}Multiclass MLP\\ + Binary SVM\end{tabular} &  38.87 & 1.40 & 69.49 & 79.87 & 38.74 & 1.24 & 68.37 & 80.01\\
    \cline{2-10}
    & Binary SVM &  \textbf{16.92} & \textbf{0.61} & \textbf{87.72} & \textbf{91.23} & \textbf{20.78} & \textbf{0.63} & \textbf{84.24} & \textbf{89.30}\\
    \cline{2-10}
    & Multiclass SVM & 27.94 & 0.17 & 82.64 & 85.94 & 35.72 & 0.18 & 76.23 & 82.05\\
    \cline{2-10}
    & Binary GBDT &  31.75 & 0.54 & 77.92 & 83.85 & 30.66 & 0.76 & 76.61 & 84.29\\
    \cline{2-10}
    & \multirow{2}{*}{Enc + MLP} &  \textbf{14.18} & \textbf{0.64} & \textbf{89.39} & \textbf{92.59} & \textbf{19.19} & \textbf{0.50} & \textbf{86.08} & \textbf{90.15} \\
    &  &   \textcolor{green}{($\downarrow$ 2.74)} & \textcolor{red}{($\uparrow$ 0.03)} & \textcolor{green}{($\uparrow$ 1.67)} & \textcolor{green}{($\uparrow$ 1.36)} & \textcolor{green}{($\downarrow$ 1.59)} & \textcolor{green}{($\downarrow$ 0.13)} & \textcolor{green}{($\uparrow$ 1.84)} & \textcolor{green}{($\uparrow$ 0.85)} \\
    \hline\hline

    \multirow{8}{*}{100}
    & Binary MLP &  20.65 & 0.46 & 86.10 & 89.44 & 23.91 & 0.44 & 83.24 & 87.82\\
    \cline{2-10}
    & Multiclass MLP &  14.44 & 4.43 & 75.42 & 90.57 & 15.81 & 4.49 & 70.34 & 89.85\\
    \cline{2-10}
    & \begin{tabular}[c]{@{}c@{}}Multiclass MLP\\ + Binary SVM\end{tabular} &  28.37 & 1.59 & 76.38 & 85.02 & 30.45 & 1.57 & 72.10 & 83.99\\
    \cline{2-10}
    & Binary SVM &  \textbf{15.06} & \textbf{0.68} & \textbf{88.56} & \textbf{92.13} & \textbf{17.91} & \textbf{0.68} & \textbf{85.49} & \textbf{90.71}\\
    \cline{2-10}
    & Multiclass SVM &  28.36 & 0.17 & 82.18 & 85.73 & 28.36 & 0.18 & 81.90 & 85.73\\
    \cline{2-10}
    & Binary GBDT &  27.76 & 0.67 & 80.15 & 85.79 & 26.62 & 0.77 & 79.20 & 86.30\\
    \cline{2-10}
    & \multirow{2}{*}{Enc + MLP} &  \textbf{12.43} & \textbf{0.53} & \textbf{90.78} & \textbf{93.52} & \textbf{16.20} & \textbf{0.45} & \textbf{88.18} & \textbf{91.68}\\
    &  &   \textcolor{green}{($\downarrow$ 2.63)} & \textcolor{green}{($\downarrow$ 0.15)} & \textcolor{green}{($\uparrow$ 2.22)} & \textcolor{green}{($\uparrow$ 1.39)} & \textcolor{green}{($\downarrow$ 1.71)} & \textcolor{green}{($\downarrow$ 0.23)} & \textcolor{green}{($\uparrow$ 2.69)} & \textcolor{green}{($\uparrow$ 0.97)}\\
    \hline\hline

    \multirow{8}{*}{200}
    & Binary MLP &  18.96 & 0.47 & 87.17 & 90.29 & 22.31 & 0.47 & 84.28 & 88.61\\
    \cline{2-10}
    & Multiclass MLP & 13.03 & 4.69 & 75.45 & 91.14 & 16.51 & 4.00 & 71.70 & 89.74\\
    \cline{2-10}
    & \begin{tabular}[c]{@{}c@{}}Multiclass MLP\\ + Binary SVM\end{tabular} &  29.93 & 1.75 & 74.48 & 84.16 & 34.50 & 1.37 & 70.38 & 82.07\\
    \cline{2-10}
    & Binary SVM & \textbf{13.76} & \textbf{0.86} & \textbf{88.64} & \textbf{92.69} & \textbf{16.55} & \textbf{0.93} & \textbf{85.16} & \textbf{91.26}\\
    \cline{2-10}
    & Multiclass SVM &  21.19 & 0.21 & 86.90 & 89.30 & 24.85 & 0.25 & 83.84 & 87.45\\
    \cline{2-10}
    & Binary GBDT &  24.71 & 0.56 & 82.71 & 87.37 & 25.24 & 0.69 & 80.83 & 87.03\\
    \cline{2-10}
    & \multirow{2}{*}{Enc + MLP} & \textbf{10.18} & \textbf{0.49} & \textbf{92.38} & \textbf{94.66} & \textbf{14.37} & \textbf{0.56} & \textbf{88.70} & \textbf{92.54}\\
    &  &   \textcolor{green}{($\downarrow$ 3.58)} & \textcolor{green}{($\downarrow$ 0.37)} & \textcolor{green}{($\uparrow$ 3.74)} & \textcolor{green}{($\uparrow$ 1.97)} & \textcolor{green}{($\downarrow$ 2.18)} & \textcolor{green}{($\downarrow$ 0.37)} & \textcolor{green}{($\uparrow$ 3.54)} & \textcolor{green}{($\uparrow$ 1.28)}\\
    \hline\hline

    \multirow{8}{*}{400}
    & Binary MLP &  \textbf{14.49} & \textbf{0.62} & \textbf{89.22} & \textbf{92.44} & \textbf{20.63} & \textbf{0.47} & \textbf{85.40} & \textbf{89.45}\\
    \cline{2-10}
    & Multiclass MLP &  14.87 & 3.45 & 77.98 & 90.84 & 17.58 & 3.37 & 73.55 & 89.53\\
    \cline{2-10}
    & \begin{tabular}[c]{@{}c@{}}Multiclass MLP\\ + Binary SVM\end{tabular} &  24.48 & 1.88 & 77.86 & 86.82 & 28.20 & 2.01 & 72.22 & 84.90\\
    \cline{2-10}
    & Binary SVM &  12.86 & 0.90 & 89.02 & 93.12 & 15.69 & 1.06 & 85.35 & 91.63\\
    \cline{2-10}
    & Multiclass SVM &  17.87 & 0.24 & 88.88 & 90.95 & 22.75 & 0.30 & 84.87 & 88.47\\
    \cline{2-10}
    & Binary GBDT &  20.16 & 0.46 & 86.24 & 89.69 & 21.91 & 0.66 & 83.40 & 88.71\\
    \cline{2-10}
    & \multirow{2}{*}{Enc + MLP} &  \textbf{8.67} & \textbf{0.44} & \textbf{93.33} & \textbf{95.44} & \textbf{14.21} & \textbf{0.59} & \textbf{88.67} & \textbf{92.60}\\
    &  &   \textcolor{green}{($\downarrow$ 5.82)} & \textcolor{green}{($\downarrow$ 0.18)} & \textcolor{green}{($\uparrow$ 4.11)} & \textcolor{green}{($\uparrow$ 3.00)} & \textcolor{green}{($\downarrow$ 6.42)} & \textcolor{red}{($\uparrow$ 0.12)} & \textcolor{green}{($\uparrow$ 3.27)} & \textcolor{green}{($\uparrow$ 3.15)}\\

    \bottomrule[1.6pt]
   \end{tabular}
 \end{table*}


\begin{table*}[h]
  \centering
  \caption{Results Before and After Removing Leak from Test Data  in Chen et al.'s Work -- AndroZoo Dataset}
  \label{tab:eval_diff_androzoo}
  \begin{tabular}{c|c|cccc||cccc}
    \toprule[1.6pt]
    \multirow{3}{*}{\begin{tabular}[c]{@{}c@{}}\textbf{Monthly}\\ \textbf{Sample}\\ \textbf{Budget}\end{tabular}} & \multirow{3}{*}{\begin{tabular}[c]{@{}c@{}}\textbf{Model}\\ \textbf{Architecture}\end{tabular}} & \multicolumn{4}{c||}{\multirow{2}{*}{\begin{tabular}[c]{@{}c@{}}\textbf{Measurement on Original Data}\\ Average Performance (\%)\end{tabular}}} & \multicolumn{4}{c}{\multirow{2}{*}{\begin{tabular}[c]{@{}c@{}}\textbf{Measurement after Leak Removed from Test}\\ Average Performance (\%)\end{tabular}}}\\                                               & &  \multicolumn{4}{c||}{} & \multicolumn{4}{c}{} \\
    & &  FNR & FPR & F1 & BA & FNR & FPR & F1 & BA \\
    \hline\hline

    \multirow{8}{*}{50}
    & Binary MLP & 51.55 & 0.30 & 60.92 & 74.08 & 58.82 & 0.40 & 53.40 & 70.39\\
    \cline{2-10}
    & Multiclass MLP &  54.84 & 15.29 & 33.45 & 64.94 & 37.82 & 32.78 & 31.09 & 64.70\\
    \cline{2-10}
    & \begin{tabular}[c]{@{}c@{}}Multiclass MLP\\ + Binary SVM\end{tabular} &  72.55 & 2.53 & 34.80 & 62.46 & 83.58 & 1.49 & 22.92 & 57.46\\
    \cline{2-10}
    & Binary SVM & \textbf{48.77} & \textbf{0.29} & \textbf{63.42} & \textbf{75.47} & \textbf{57.56} & \textbf{0.28} & \textbf{54.28} & \textbf{71.08}\\
    \cline{2-10}
    & Multiclass SVM & 64.96 & 0.12 & 47.89 & 67.46 & 72.50 & 0.09 & 39.16 & 63.70\\
    \cline{2-10}
    & Binary GBDT &  50.35 & 0.47 & 61.06 & 74.59 & 58.34 & 0.48 & 52.48 & 70.59\\
    \cline{2-10}
    & \multirow{2}{*}{Enc + MLP} & \textbf{29.79} & \textbf{0.53} & \textbf{79.29} & \textbf{84.84} & \textbf{55.65} & \textbf{0.47} & \textbf{56.83} & \textbf{71.94}\\
    &  &   \textcolor{green}{($\downarrow$ 18.98)} & \textcolor{red}{($\uparrow$ 0.24)} & \textcolor{green}{($\uparrow$ 15.87)} & \textcolor{green}{($\uparrow$ 9.37)} & \textcolor{green}{($\downarrow$ 1.91)} & \textcolor{red}{($\uparrow$ 0.19)} & \textcolor{green}{($\uparrow$ 2.55)} & \textcolor{green}{($\uparrow$ 0.86)}\\
    \hline\hline

    \multirow{8}{*}{100}
    & Binary MLP & 45.79 & 0.31 & 65.80 & 76.95 & 58.22 & 0.29 & 54.08 & 70.75\\
    \cline{2-10}
    & Multiclass MLP &  42.23 & 21.24 & 36.59 & 68.27 & 56.76 & 20.54 & 27.61 & 61.35\\
    \cline{2-10}
    & \begin{tabular}[c]{@{}c@{}}Multiclass MLP\\ + Binary SVM\end{tabular} & 77.32 & 1.59 & 31.51 & 60.55 & 80.44 & 2.01 & 25.78 & 58.77\\
    \cline{2-10}
    & Binary SVM &  \textbf{43.07} & \textbf{0.32} & \textbf{68.33} & \textbf{78.31} & \textbf{52.28} & \textbf{0.29} & \textbf{59.23} & \textbf{73.72}\\
    \cline{2-10}
    & Multiclass SVM &  53.97 & 0.06 & 59.14 & 72.99 & 64.04 & 0.10 & 47.90 & 67.93\\
    \cline{2-10}
    & Binary GBDT & 48.59 & 0.76 & 62.58 & 75.33 & 57.56 & 0.65 & 53.62 & 70.90\\
    \cline{2-10}
    & \multirow{2}{*}{Enc + MLP} &  \textbf{30.17} & \textbf{0.44} & \textbf{79.46} & \textbf{84.70} & \textbf{47.19} & \textbf{0.37} & \textbf{64.94} & \textbf{76.22} \\
    &  &   \textcolor{green}{($\downarrow$ 12.90)} & \textcolor{red}{($\uparrow$ 0.12)} & \textcolor{green}{($\uparrow$ 11.13)} & \textcolor{green}{($\uparrow$ 6.39)} & \textcolor{green}{($\downarrow$ 5.09)} & \textcolor{red}{($\uparrow$ 0.08)} & \textcolor{green}{($\uparrow$ 5.71)} & \textcolor{green}{($\uparrow$ 2.50)}\\
    \hline\hline

    \multirow{8}{*}{200}
    & Binary MLP &  43.58 & 0.27 & 67.63 & 78.07 & 56.16 & 0.22 & 56.12 & 71.81\\
    \cline{2-10}
    & Multiclass MLP &  32.42 & 24.78 & 37.74 & 71.40 & 46.85 & 35.64 & 26.90 & 58.76\\
    \cline{2-10}
    & \begin{tabular}[c]{@{}c@{}}Multiclass MLP\\ + Binary SVM\end{tabular} &  65.71 & 1.75 & 42.32 & 66.27 & 75.32 & 0.90 & 33.92 & 61.89\\
    \cline{2-10}
    & Binary SVM & \textbf{40.31} & \textbf{0.37} & \textbf{70.24} & \textbf{79.66} & \textbf{48.94} & \textbf{0.37} & \textbf{61.80} & \textbf{75.34}\\
    \cline{2-10}
    & Multiclass SVM &  46.55 & 0.16 & 65.56 & 76.64 & 59.44 & 0.14 & 52.11 & 70.21\\
    \cline{2-10}
    & Binary GBDT &  42.97 & 0.80 & 67.28 & 78.12 & 53.07 & 0.72 & 58.28 & 73.11\\
    \cline{2-10}
    & \multirow{2}{*}{Enc + MLP} &  \textbf{29.40} & \textbf{0.34} & \textbf{79.35} & \textbf{85.13} & \textbf{43.21} & \textbf{0.43} & \textbf{68.08} & \textbf{78.18}\\
    &  &   \textcolor{green}{($\downarrow$ 10.91)} & \textcolor{red}{($\uparrow$ 0.03)} & \textcolor{green}{($\uparrow$ 9.11)} & \textcolor{green}{($\uparrow$ 5.47)} & \textcolor{green}{($\downarrow$ 5.73)} & \textcolor{red}{($\uparrow$ 0.06)} & \textcolor{green}{($\uparrow$ 6.28)} & \textcolor{green}{($\uparrow$ 2.84)}\\
    \hline\hline

    \multirow{8}{*}{400}
    & Binary MLP &  35.96 & 0.41 & 73.99 & 81.81 & \textbf{42.33} & \textbf{0.24} & \textbf{69.55} & \textbf{78.72}\\
    \cline{2-10}
    & Multiclass MLP & 24.98 & 29.96 & 37.56 & 72.53 & 36.28 & 38.09 & 30.23 & 62.82\\
    \cline{2-10}
    & \begin{tabular}[c]{@{}c@{}}Multiclass MLP\\ + Binary SVM\end{tabular} &  61.55 & 1.67 & 46.81 & 68.39 & 81.04 & 1.29 & 27.14 & 58.84\\
    \cline{2-10}
    & Binary SVM &  34.73 & 0.43 & 74.12 & 82.42 & 43.18 & 0.46 & 66.39 & 78.18\\
    \cline{2-10}
    & Multiclass SVM &  41.75 & 0.13 & 69.59 & 79.06 & 51.80 & 0.08 & 59.54 & 74.06\\
    \cline{2-10}
    & Binary GBDT &  \textbf{33.62} & \textbf{0.38} & \textbf{76.82} & \textbf{83.00} & 42.95 & 0.39 & 69.12 & 78.33\\
    \cline{2-10}
    & \multirow{2}{*}{Enc + MLP} &  \textbf{23.59} & \textbf{0.37} & \textbf{84.23} & \textbf{88.02} & \textbf{35.51} & \textbf{0.52} & \textbf{73.73} & \textbf{81.98}\\
    &  &   \textcolor{green}{($\downarrow$ 10.03)} & \textcolor{green}{($\downarrow$ 0.01)} & \textcolor{green}{($\uparrow$ 7.41)} & \textcolor{green}{($\uparrow$ 5.02)} & \textcolor{green}{($\downarrow$ 6.82)} & \textcolor{red}{($\uparrow$ 0.28)} & \textcolor{green}{($\uparrow$ 4.18)} & \textcolor{green}{($\uparrow$ 3.26)}\\

    \bottomrule[1.6pt]
   \end{tabular}
\end{table*}


We examined how the presence of train-test leakage might impact the
main research conclusion drawn by the authors. We did so by evaluating
the models both on the original test data, and on the test data after
removing the leakage.  We then compared the results.  We expected that
all models' performances would drop after leakage in test data is
removed. The key question is whether the main conclusion in the paper,
that the proposed continuous learning approach outperforms the prior
approaches, will still hold after leakage is removed from test data.
Before we conducted the experiments, we first ran the model code on
the datasets published by the authors to see whether we could
replicate similar performance results as reported in Chen et al.'s
original paper~\cite{Chen:USENIX23}.  Apart from small differences in
the performance numbers due to differences in numerical precision
across various hardware architectures and non-deterministic operations
related to parallelism and threading\footnote{ Due to these factors
the model might not produce exactly the same results even when run on
the same data with the same settings twice.}, our results replicated
those reported in the original paper.

We then re-ran all the experiments to obtain the new performance
measurement on testing data with leakage removed.  We kept the rest of
the experiments and setup the same as in the original
paper~\cite{Chen:USENIX23}. Tables~\ref{tab:eval_diff_apigraph} and
\ref{tab:eval_diff_androzoo} show the performance results of the
various models before and after the train-test leakage is removed.  We
follow the same table format as that used in the original
paper~\cite{Chen:USENIX23}, and add a set of columns for performance
metrics measured on the test data when the leakage has been removed.
The original paper used FNR, FPR, and F1 score. We also included
Balanced Accuracy (BA) as an additional metric.  In both tables, all
columns under ``Measurement on Original Data'' are produced by our own
experiments replicating the results from the original paper.  All
columns under ``Measurement after Leak Removed from Test'' are results
obtained after we remove leakage from the testing apps before
measuring the models' performance.  The table is organized by blocks
for monthly sample budget from 50 to 400. For a fixed monthly sample
budget, the authors' approach (\mbox{Enc + MLP}) is compared with the
best performance from the prior approaches, both indicated in bold
font. The value in the colored parentheses with up or down arrows
indicate the difference between the \mbox{Enc + MLP} approach and the
best prior approaches, with green in favor of \mbox{Enc + MLP} and red
in favor of the best prior approach.

For both APIGraph and AndroZoo datasets, there is a consistent
decrease of the overall model performance when leakage is removed from
test data.  The F1 metrics overall show decrease after leakage is
removed from testing data. But F1 metric is also sensitive to the
class ratio in the testing data. If the leakage removal reduces the
malware ratio in the testing data, this would result in a decreased F1
score even if the model's predicting capability remains the same.  As
discussed in Section~\ref{sec:metrics}, the BA metric isolates the
performance results from the change in the malware ratio in the test
set.  By comparing the BA numbers before and after leakage removal,
one can see how much performance is inflated by the train-test
leakage.  We can see that the two BA metric columns for the same row
almost always show a decrease when leak is removed from test. This
decrease is more notable for the AndroZoo data
(Table~\ref{tab:eval_diff_androzoo}) than for the APIGraph data
(Table~\ref{tab:eval_diff_apigraph}). This confirms our hypothesis
that train-test leakage inflates a machine learning model's measured
performance.

We now examine whether the inflated performance results due to
train-test leakage would impact the overall research conclusion, in
this case whether \mbox{Enc+MLP} outperforms the prior approaches. We
notice that with only one exception (FPR number of monthly labeling
budget 400 in both tables) the color-coded comparison results remain
the same color before and after leakage removal. If we only look at
the comparative results for the overall metrics of F1 and BA, the
color-coded differences all show the same green color, meaning the
\mbox{Enc+MLP} outperforms the best from the prior baseline
approaches. Thus, while the train-test leakage inflated the
performance results of all models, the relative performance between
\mbox{Enc+MLP} and the prior models, as measured by F1 score and BA,
remain the same.  {\it The overall conclusion from Chen et al.'s work
still holds.} We do notice, however, that for the AndroZoo
data~(Table~\ref{tab:eval_diff_androzoo}), the \mbox{Enc+MLP}
approach's improvement margin from best prior works considerably
shrinks after leakage is removed from test data.

 \section{Case Study 2: Graph Convolutional Networks for Android Malware Detection}
\label{sec:case_ours}

Our second case study is research of the authors of this paper. 
We have  investigated the use of graph
convolutional networks (GCN) for Android malware detection. We apply
GCN on Android apps' interprocedural control flow graphs (ICFG) to build a
deep learning model for Android malware detection, and compare the performance of the 
GCN model against that of random forest, a commonly used
traditional ML model. In this section, we report on the train-test
leakage identified in this research of our own, and how it impacts the
conclusion we could draw from the research.

\subsection{Summary of the Research and Dataset Used}
\label{sec:summary_ours}

\begin{figure*}[h]
     \centering
     \begin{subfigure}{0.49\textwidth}
         \centering
         \includegraphics[width=\textwidth]{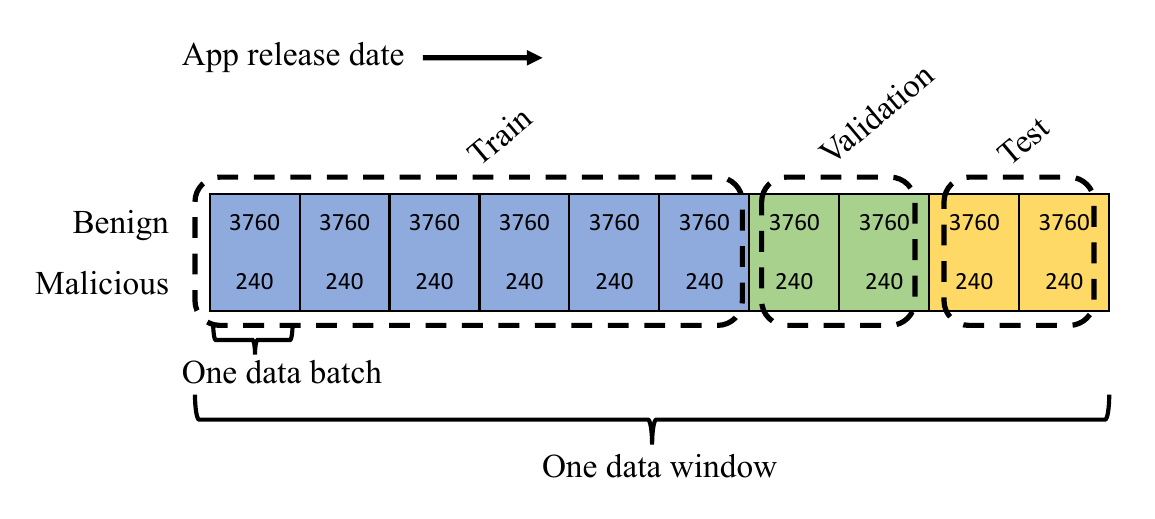}
         \caption{Breakdown of a sliding window}
         \label{fig:window_detail}
     \end{subfigure}
     \hfill
     \begin{subfigure}{0.49\textwidth}
         \centering
         \includegraphics[width=\textwidth]{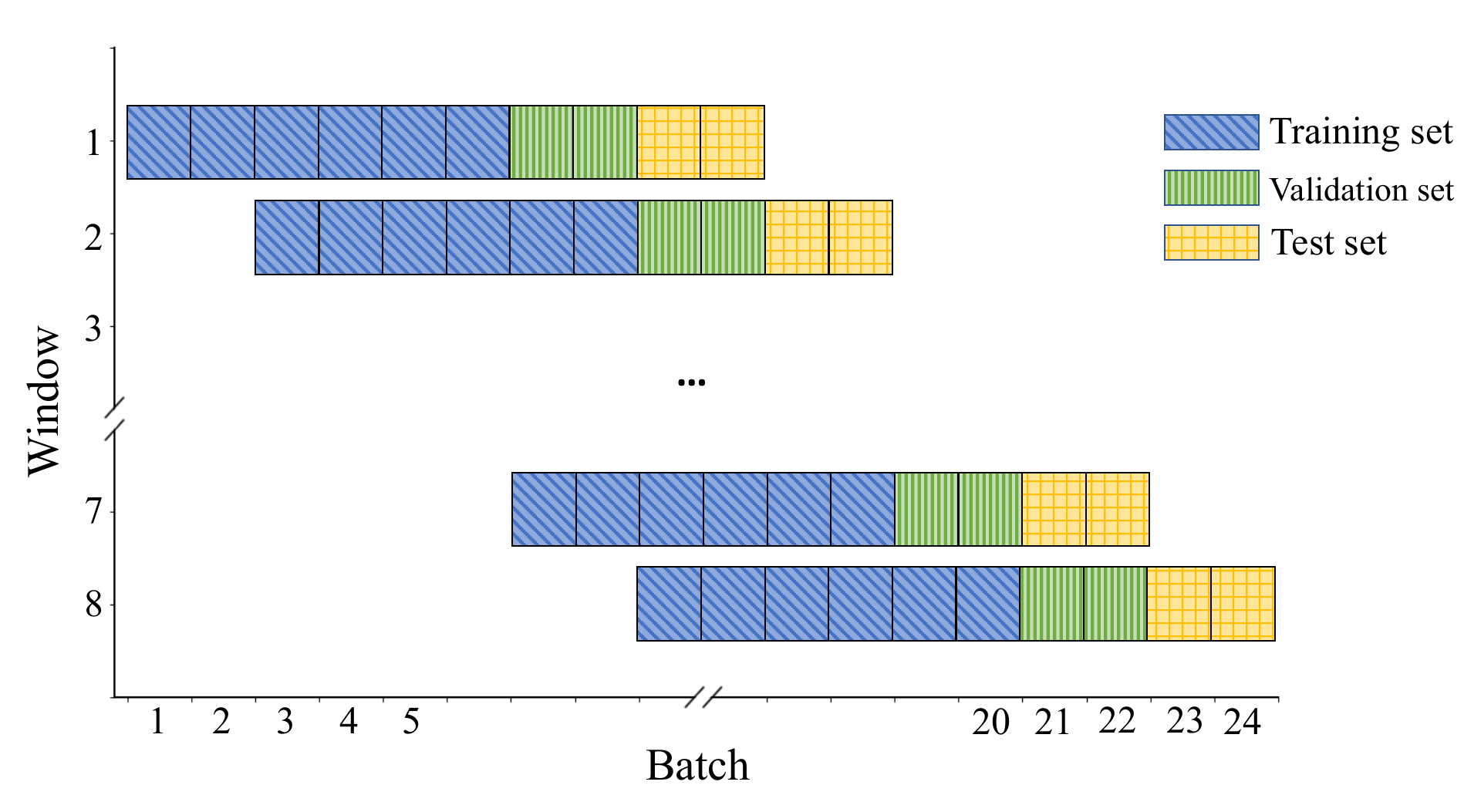}
         \caption{Data window sliding}
         \label{fig:sliding_window_overview}
       \end{subfigure}
       \caption{Data Sliding Windows}
       \label{fig:data_sliding_windows}
 \end{figure*}

GCNs leverage graph structures as input, which allows for
potentially capturing richer static code semantics, such as that expressed in
an Android app's ICFG.
We use an Android app's ICFG\footnote{
To obtain a single ICFG for an Android app, we combine the ICFG of each component of the app based on the life cycle semantics of how the methods of various components are called by the
Android runtime.}
as the GCN model's input, where each node represents an API call,
with additional attributes including class path, method name,
parameter type, and node degree. Following the standard GCN training
process, we obtain a graph embedding for the ICFG, 
achieved through multiple rounds of message passing within
the GCN, where each node’s embedding is updated based on its
neighbors.  Finally, a readout layer aggregates these node embeddings
into a unified graph embedding for the entire ICFG. This component of the model can be seen as a graph encoder. The unified graph embedding learned by the encoder is then fed as input to the classification layer of the model. More details
about the GCN model can be found in the Appendix B.

We compare the GCN approach to a standard ML approach -- random
forest (RF). The RF model takes as input a feature vector representing
an app. We extract DREBIN-like static features and rank them in decreasing order of
their mutual information with the class label. We then select 2,970
most informative
features for training an RF model, which include permissions, intent
actions, discriminative APIs, obfuscation signatures, and native code
signatures.
The mutual information is close to 0 for features ranked lower.

We collected Google Play apps released between January 2018 and
December 2021 through AndroZoo~\cite{Allix:MSR16}, and used
VirusTotal~\cite{VirusTotal} results to label the apps.
Using apps from a single market prevents the sampling bias~\cite{Arp:USENIXSeurity22}
as mentioned in Section~\ref{sec:biases}.
We follow a sliding window approach to construct training, validation,
and testing datasets, as shown in Figure~\ref{fig:data_sliding_windows}.
We first split malicious app data into batches along app release
date, each batch initially consisting of 240 malicious apps.  
Then, we randomly sample benign apps from all the benign apps in a
batch's time frame to achieve a $6\%$ malware ratio to mimic real
malware distribution. Consequently, each batch comprises 240 malicious
apps and 3,760 benign apps from the corresponding timeframe, for a
total of 4,000 apps in a batch.  This sampling based on release date
avoids the spatial bias~\cite{Pendlebury:USENIX19} as discussed in
Section~\ref{sec:biases}.  We subsequently form windows of ten
consecutive batches, with the first six batches used for training, the
next two for validation, and the last two for testing.

The complete dataset encompasses 24 batches, totaling 5,760 malicious and 90,240
benign apps, all chronologically ordered by their release date on Google Play.
This dataset provides eight sliding windows (the window slides two batches each time)
for evaluating machine learning models through different training, validation,
and testing data. This sliding window approach, as the  sampling approach for  batch construction, also avoids the temporal bias~\cite{Pendlebury:USENIX19}
as discussed in Section~\ref{sec:biases}.

\subsection{Identification of Train-Test Leakage}
\label{sec:leak_ours}

 \begin{figure}[h]
     \centering
     \includegraphics[width=0.45\textwidth]{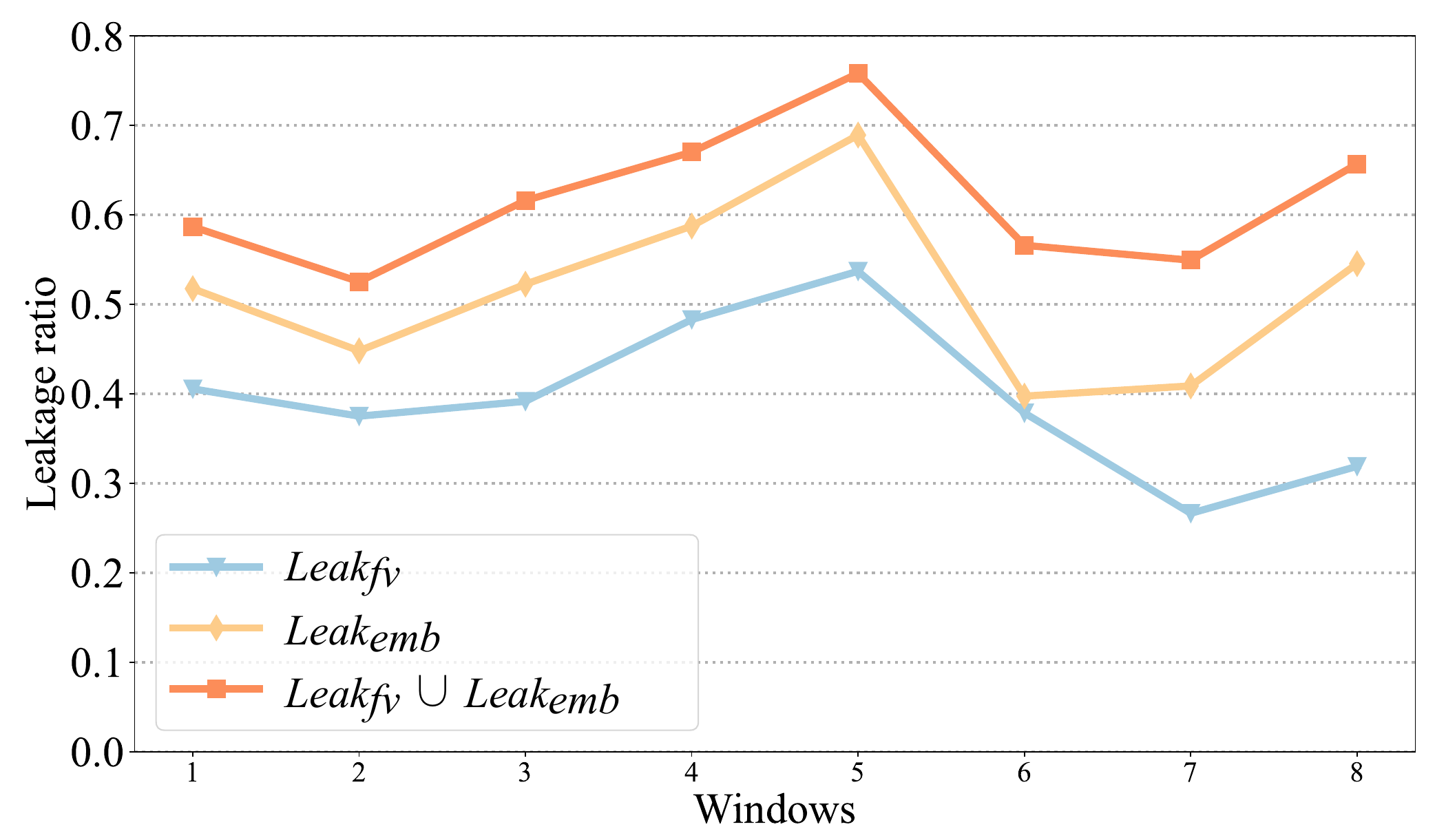}
     \caption{Ratio of leakage apps in sliding-window-based test data (2018-01 to 2021-12)}
     \label{fig:window_identical_vec}
 \end{figure}

The two machine learning models our research compares, GCN and RF,
use different input representations for an Android app. In GCN, an input app
is represented as an ICFG adjacency matrix, together with a list of  attributes for each node, whereas in RF an input app is
represented as a feature vector. In prior discussion, we have
explained the definition of train-test leakage when 
a feature vector is used as an app's representation.
These are apps in the testing data that share
identical feature vectors with at least one app in the training
data. We denote these leakage apps as  \textbf{$Leak_{\textit{fv}}$}.
While this definition  works well for RF, it is not suitable to use in identifying train-test leakage for the GCN-based
models since there an app is represented as a graph. 
Given that there is currently no known efficient algorithm to solve
graph isomorphism, it is not clear how to
efficiently determine if two Android
apps' ICFGs are identical. 
However, as mentioned above, the GCN model consists of an encoder and a classification layer. The encoder generates a graph embedding, i.e., a vector of floating numbers, which is provided as input to the classification layer. 

Instead of comparing the ICFGs directly, we
have chosen to use the computed graph embeddings of the ICFGs
to identify train-test leakage. Specifically,
given the  embeddings $emb_{1}$ and $emb_{2}$ of two Android apps' ICFGs,  we
compute cosine similarity between the two embeddings as 

\[ \textit{cosine similarity} = \frac{emb_{1} \cdot emb_{2}}{\|emb_{1}\|\|emb_{2}\|} \]

\noindent
where $\cdot$ is the inner product operation and $\| v \|$ represents the length of a vector $v$.
The cosine similarity measures how similar the two vectors are and ranges from $0$ to $1$,
where $1$ means the two vectors are identical, while $0$ means that the two vectors are orthogonal and share no similarity.

Note that due to GCN model's stocasticity and the numerical precision limitations, even if we feed the exact same ICFG multiple times to a GCN, the model will not generally produce the exact same graph embeddings. Thus, for the purpose of identifying train-test leakage, we chose
a threshold $M$. A test app is considered a leakage if there is at least one training app
whose cosine similarity with the testing app is at least $M$.
We denote these apps as \textbf{$Leak_{\textit{emb}}$}. For a given testing set,
we choose $M$ so that the two sets of leakage apps in the testing data,
$Leak_{\textit{fv}}$ and $Leak_{\textit{emb}}$, are most aligned. The $M$ value
varies from one testing window (Figure~\ref{fig:data_sliding_windows}) to another,
but they are all very close to $1$ (e.g., $0.9998$). More details on how the threshold
$M$ is determined can be found in Appendix C.

Figure~\ref{fig:window_identical_vec} illustrates the ratio of
leakage apps ($Leak_{\textit{fv}}$, $Leak_{\textit{emb}}$, and their union) on the
testing data used in our experiments. 
A significant portion of apps in the testing data (oftentimes exceeding half of the total apps)
are identified as train-test leakage. 
Since the two models give rise to two different sets of leakage apps,
we need to remove the union of the two leakage sets from the testing data in order
to compare the two models' performance after leakage removal,
for the same reasons as explained in Section~\ref{sec:leak_chen}.

\subsection{Impact of the Leakage on Performance Results}
\label{sec:experiment_ours}

\begin{table*}[h]
  \centering
  \caption{Compare GCN and RF Before and After Leakage Removal}
  \label{tab:gcn_rf_results}
   \begin{tabular}{l|c|cccc||cccc}
  \toprule[1.6pt]
     \multirow{3}{*}{\textbf{Window}} & \multirow{3}{*}{\begin{tabular}[c]{@{}c@{}}\textbf{Model}\\ \textbf{Architecture}\end{tabular}}  & \multicolumn{4}{c||}{\multirow{2}{*}{\begin{tabular}[c]{@{}c@{}}\textbf{Measurement on Complete Test Data}\\ Performance (\%)\end{tabular}}} & \multicolumn{4}{c}{\multirow{2}{*}{\begin{tabular}[c]{@{}c@{}}\textbf{Measurement after Leak Removed from Test} \\ Performance (\%)\end{tabular}}}\\
  & &  \multicolumn{4}{c||}{} & \multicolumn{4}{c}{} \\
  &  & FNP & FPR & F1  & BA  & FNP & FPR & F1  & BA \\
  \hline\hline
  \multirow{3}{*}{Window\_1}
   & RF & 15.83 & 0.01 & 91.30 & 92.08 & 72.83 & 0.03 & 42.37 & 63.57 \\ 
   \cline{2-10}
   & \multirow{2}{*}{GCN} & 16.88 & 2.25 & 76.15 & 90.44 & 40.22 & 2.18 & 50.69 & 78.80 \\ 
   &  & \textcolor{red}{($\downarrow$1.05)} & \textcolor{red}{($\downarrow$2.24)} & \textcolor{red}{($\downarrow$15.15)} & \textcolor{red}{($\downarrow$1.64)}  & \textcolor{green}{($\uparrow$32.61)} & \textcolor{red}{($\downarrow$2.15)} & \textcolor{green}{($\uparrow$8.32)} & \textcolor{green}{($\uparrow$15.27)} \\
   \hline\hline

   \multirow{3}{*}{Window\_2}
   & RF & 12.08 & 0.01 & 93.47 & 93.95 & 69.51 & 0.03 & 46.30 & 65.23 \\ 
   \cline{2-10}
   & \multirow{2}{*}{GCN} & 14.79 & 0.64 & 87.30 & 92.29 & 57.32 & 0.13 & 57.38 & 71.27 \\ 
   &  & \textcolor{red}{($\downarrow$2.71)} & \textcolor{red}{($\downarrow$0.63)} & \textcolor{red}{($\downarrow$6.17)} & \textcolor{red}{($\downarrow$1.66)} & \textcolor{green}{($\uparrow$12.19)} & \textcolor{red}{($\downarrow$0.10)} & \textcolor{green}{($\uparrow$11.08)} & \textcolor{green}{($\uparrow$6.04)} \\
   \hline\hline

  \multirow{3}{*}{Window\_3}
   & RF & 12.29 & 0.15 & 92.32 & 93.78 & 68.89 & 0.44 & 42.75 & 65.34 \\ 
   \cline{2-10}
   & \multirow{2}{*}{GCN} & 17.92 & 1.53 & 79.68 & 90.28 & 64.44 & 0.30 & 48.85 & 67.63 \\ 
   &  & \textcolor{red}{($\downarrow$5.63)} & \textcolor{red}{($\downarrow$1.38)} & \textcolor{red}{($\downarrow$12.64)} & \textcolor{red}{($\downarrow$3.50)} & \textcolor{green}{($\uparrow$4.45)} & \textcolor{green}{($\uparrow$0.14)} & \textcolor{green}{($\uparrow$6.10)} & \textcolor{green}{($\uparrow$2.29)} \\
   \hline\hline

  \multirow{3}{*}{Window\_4}
   & RF & 11.25 & 0.37 & 91.22 & 94.19 & 51.38 & 0.00 & 65.43 & 74.31 \\ 
   \cline{2-10}
   & \multirow{2}{*}{GCN} & 18.96 & 2.54 & 73.40 & 89.25 & 42.20 & 0.40 & 69.23 & 78.70 \\ 
   &  & \textcolor{red}{($\downarrow$7.71)} & \textcolor{red}{($\downarrow$2.17)} & \textcolor{red}{($\downarrow$17.82)} & \textcolor{red}{($\downarrow$4.94)} & \textcolor{green}{($\uparrow$9.18)} & \textcolor{red}{($\downarrow$0.40)} & \textcolor{green}{($\uparrow$3.80)} & \textcolor{green}{($\uparrow$4.39)} \\
   \hline\hline

  \multirow{3}{*}{Window\_5}
   & RF & 13.54 & 0.86 & 86.46 & 92.80 & 64.89 & 0.61 & 48.94 & 67.25 \\ 
   \cline{2-10}
   & \multirow{2}{*}{GCN} & 15.62 & 3.26 & 71.68 & 90.56 & 39.69 & 0.72 & 70.85 & 79.79 \\ 
   &  & \textcolor{red}{($\downarrow$2.08)} & \textcolor{red}{($\downarrow$2.40)} & \textcolor{red}{($\downarrow$14.78)} & \textcolor{red}{($\downarrow$2.24)} & \textcolor{green}{($\uparrow$25.20)} & \textcolor{red}{($\downarrow$0.11)} & \textcolor{green}{($\uparrow$21.91)} & \textcolor{green}{($\uparrow$12.54)} \\
   \hline\hline

  \multirow{3}{*}{Window\_6}
   & RF & 22.71 & 0.17 & 85.88 & 88.56 & 84.93 & 0.21 & 25.14 & 57.43 \\ 
   \cline{2-10}
   & \multirow{2}{*}{GCN} & 22.50 & 2.45 & 71.81 & 87.53 & 43.15 & 0.66 & 66.14 & 78.09 \\ 
   &  & \textcolor{red}{($\downarrow$0.21)} & \textcolor{red}{($\downarrow$2.28)} & \textcolor{red}{($\downarrow$14.07)} & \textcolor{red}{($\downarrow$1.03)} & \textcolor{green}{($\uparrow$41.78)} & \textcolor{red}{($\downarrow$0.45)} & \textcolor{green}{($\uparrow$41.00)} & \textcolor{green}{($\uparrow$20.66)} \\
   \hline\hline

  \multirow{3}{*}{Window\_7}
   & RF & 9.79 & 3.03 & 75.90 & 93.59 & 16.58 & 0.25 & 89.91 & 91.59 \\ 
   \cline{2-10}
   & \multirow{2}{*}{GCN} & 17.92 & 2.57 & 73.85 & 89.76 & 18.98 & 1.45 & 83.70 & 89.78 \\ 
   &  & \textcolor{red}{($\downarrow$8.13)} & \textcolor{green}{($\uparrow$0.46)} & \textcolor{red}{($\downarrow$2.05)} & \textcolor{red}{($\downarrow$3.83)} & \textcolor{red}{($\downarrow$2.40)} & \textcolor{red}{($\downarrow$1.20)} & \textcolor{red}{($\downarrow$6.21)} & \textcolor{red}{($\downarrow$1.81)} \\
   \hline\hline

  \multirow{3}{*}{Window\_8}
   & RF & 15.00 & 1.10 & 84.04 & 91.95 & 84.95 & 0.15 & 25.23 & 57.45 \\ 
   \cline{2-10}
   & \multirow{2}{*}{GCN} & 19.79 & 1.76 & 77.23 & 89.23 & 54.84 & 0.49 & 56.76 & 72.34 \\ 
   &  & \textcolor{red}{($\downarrow$4.79)} & \textcolor{red}{($\downarrow$0.66)} & \textcolor{red}{($\downarrow$6.81)} & \textcolor{red}{($\downarrow$2.72)} & \textcolor{green}{($\uparrow$30.11)} & \textcolor{red}{($\downarrow$0.34)} & \textcolor{green}{($\uparrow$31.53)} & \textcolor{green}{($\uparrow$14.89)} \\

   \bottomrule[1.6pt]
   \end{tabular}
\end{table*}


 Like in Section~\ref{sec:case_chen}, we compare the performance measurements of the
 two models both before and after leakage apps are removed from testing data.
 Leakage portion is defined as $Leak_{\textit{fv}} \cup Leak_{\textit{emb}}$.
Table~\ref{tab:gcn_rf_results} shows the results. For each testing window, the color-coded performance
differences are included in parentheses, with green in favor of GCN and red in favor
of RF.
If we measure the performance on all the test data, RF consistently outperforms GCN
based on the F1 score and balanced accuracy metric. {\it However, once the leakage is removed,
the comparative result flips:} GCN outperforms RF in seven out of the eight testing
windows, and by big margins.  It is clear that after leakage is removed, not only both models' performance
decreases, but their comparative results also change to the opposite. In this case,
{\it the train-test leakage would have led to a completely different conclusion being
drawn about the comparative advantage between the GCN and RF models on Android malware
detection.}

Therefore, train-test leakage can lead to two key issues: inflated research performance and misleading research conclusions. The inflated performance occurs because the model is inadvertently trained on information it will encounter during testing. This can skew the results and make the model appear more effective than it truly is. Consequently, these misleading results can influence research conclusions.

\section{Which is Important: Memorization or Generalization?}
\label{sec:practical_impact_ours}

In Section~\ref{sec:intro} we set out to answer two questions:
1) how will train-test leakage impact an ML model's measured performance
for Android malware detection,
and 2) given that such leakage will happen ``naturally'' when the
model is applied to real-world data, is there any practical value
in discerning such differences of performance measurement?
At this point we have clearly answered question 1). 
In this
section we look at question 2).
In our experiments we remove the leakage apps from testing data before evaluating
a model's performance. These leakage apps
would be present in the real world data when the ML model is applied; wouldn't evaluation
on the complete test data (including the leakage apps) reflect the model's utility
in the real world? This question is especially relevant when research conclusion
about different models' comparative advantages flips when leakage data is removed (e.g., in case study 2).
Should practitioners care about this result, if all they need is a model
that would perform better on the real world data?

\begin{figure}[t]
     \centering
     \begin{subfigure}{0.45\textwidth}

       \includegraphics[width=\textwidth]{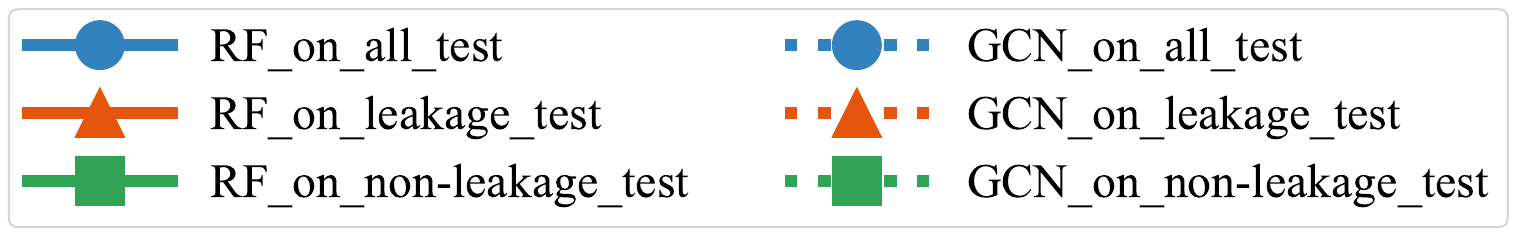}
         \label{fig:leakage_legend}
         \hspace{-2.0em}
     \end{subfigure}
 
     \begin{subfigure}{0.49\textwidth}
         \centering
         \includegraphics[width=\textwidth]{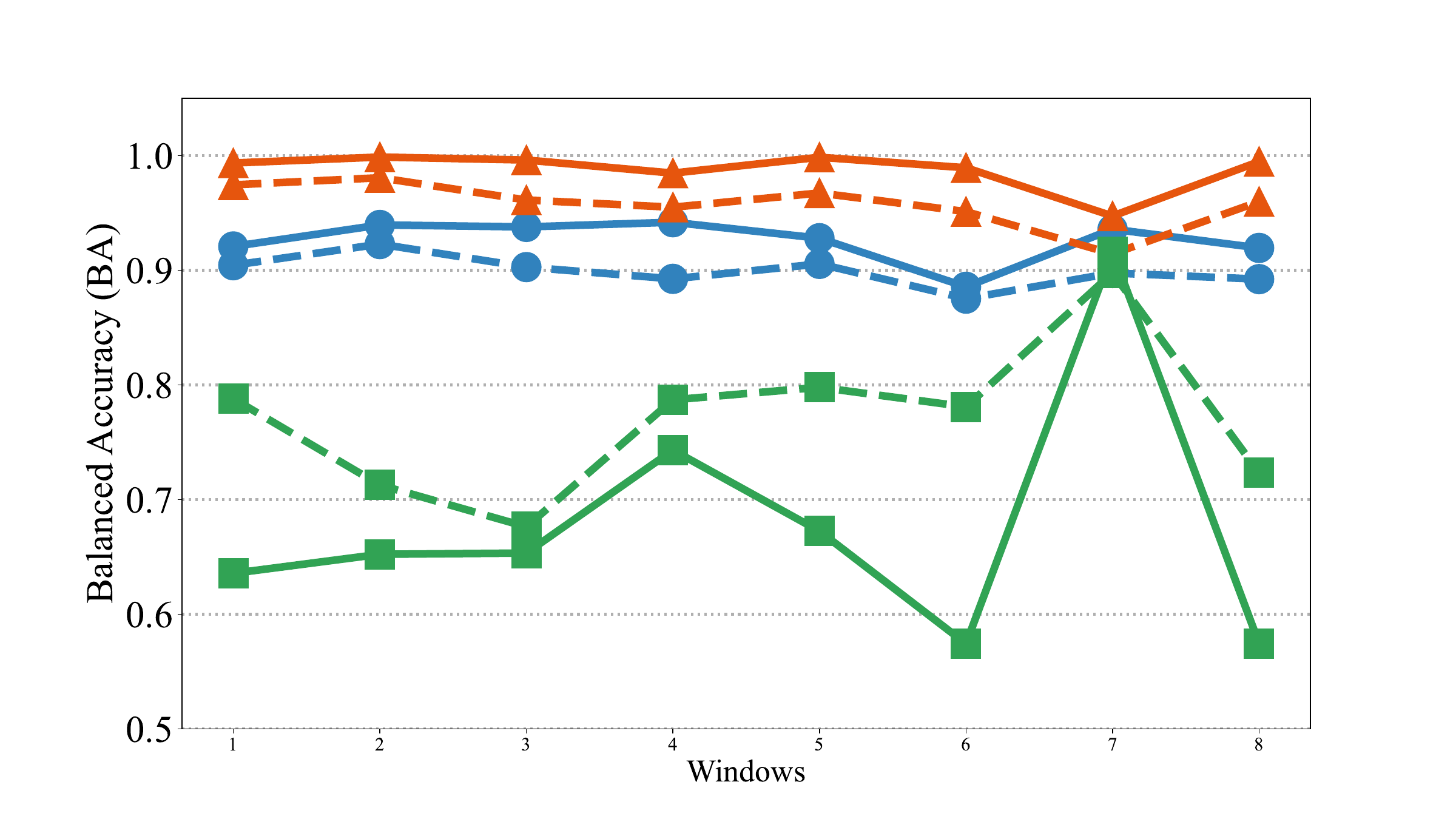}
         \label{fig:union_leakage}
     \end{subfigure}

     \caption{Performance comparison of RF and GCN on the leakage and non-leakage portions of the test data, as well as
       on the complete test data.}
      \label{Fig:leakage_union}
 \end{figure}

In real-world operation, all data must be processed. Removing the
leakage from the test set is intended to evaluate the ML model's
capacity to generalize -- to determine whether good performance is due
to the model's ability to generalize to unseen samples or simply
because it has memorized the data it has already seen. This
differentiation is important for deciding which ML model to use, or
whether to use ML at all. 
 It will be helpful to understand why the comparative results flipped
 in case study 2. 
Figure~\ref{Fig:leakage_union} illustrates the performance of RF and
GCN models on 
1) the complete test data, 2)
the leakage portion of the test data, and 3) the non-leakage portion
of the test data. 
Non-leakage portion is the test data after the leakage is removed.
The figure shows that 
both models achieve their highest performance when tested on the leakage portion of
the test data and the
lowest on the non-leakage portion of test data. But their comparative
differences are opposite between the two portions. RF outperforms GCN on
the leakage portion, indicating that RF is better than GCN at memorization.
GCN outperforms RF on non-leakage portion, indicating that GCN is better
than RF at generalization. It turns out that RF's
advantage at memorization outweights its disadvantage on generalization,
and shows stronger overall performance on the complete test data. The question
is, does this mean RF should be recommended as having better
utility for real-world operation?
In other words, which capability is more important for
the {\it practical utility} of a machine learning model:
memorization or generalization? The answer is resoundinlgy the latter.
This is because for apps that are known malware, there are
other more efficient methods to detect them, such as through
a known signature. Only when a malware has not been widely
recognized a machine learning model could potentially provide
useful guidance. Thus, evaluating the machine learning
model on the non-leakage part of the testing data is more
meaningful than evaluating on the complete testing data.

To further illustrate this point, 
we create a hypothetical detector, which we call ``leak-aware detector,''
for the purpose of approximating the effect of combining machine learning 
and signature-based approaches to detect Android malware. 
For an app the detector needs to make a decision on, we first determine if
the app is considered leakage, i.e., is there an app in the training
data that has identical or highly similar representation to it?
If so we identify all such apps and use their labels' majority vote to make a decision.
Otherwise we use the trained machine learning model to make a prediction. 
This clearly separates generalization and memorization: we only use machine
learning for generalizing, and use signature matching for memorizing.
This leak-aware scheme more accurately reflects a machine learning model's
utility in practice.
It is designed as an approximate model
for combining machine learning with signature detection and used as a measurement
tool only. It is by no means an indication on
how such combination shall be implemented for operations.
In fact it is not an efficient implementation since the detector
needs to constantly keep track of the set of apps that have been used in training
the machine learning model. 
We implemented such leak-aware detectors for the models in case study 2, to
answer the question of \textit{``whether the flipped research conclusion makes any
difference in the real world.''} For each model -- RF and GCN, we utilize
the same leakage definition -- $Leak_{\textit{fv}} \cup Leak_{\textit{emb}}$
to build the leak-aware detector described above.

\begin{figure}[t]
  \centering
       \includegraphics[width=.5\textwidth]{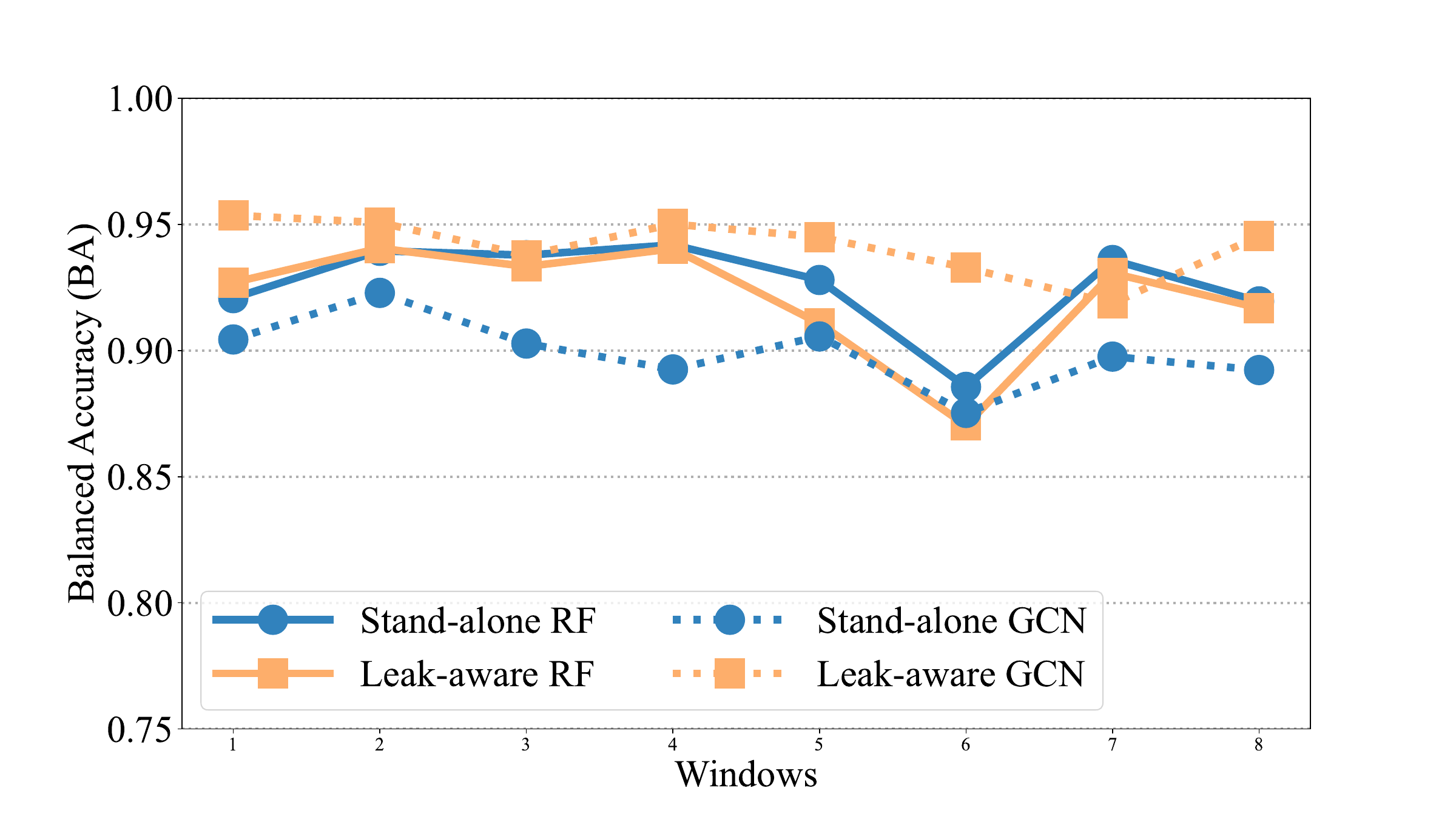}
         \caption{Leak-aware vs. Stand-alone Detector Performance}
         \label{fig:leak_aware_performance}
\end{figure}

The performance results
of the two leak-aware detectors, along with the stand-alone versions (applying
the ML model on all test data), are shown in Figure~\ref{fig:leak_aware_performance}.
The leak-aware GCN detector's overall performance on the
complete test data exceeds both the stand-alone RF and the leak-aware RF.
This is in contrast to the comparison of stand-alone GCN and stand-alone RF,
where RF outperforms GCN on the complete test data.

This result shows that not only GCN outperforms RF on the testing data
with leakage removed, it also outperforms RF on the complete test data
when used in the leak-aware scheme. Thus, the flipped comparative
result between the two models is not only relevant to understanding the models' ability to
generalize to data it has not seen, it also has practical implications
when the models are applied to real-world data and in combination with
signature-based detection schemes. 
Specifically, by approximating the effect of such combination, 
the leak-aware detector allows the GCN's comparative strength against RF to
show on the complete testing data as well.

\section{Related Work}
\label{sec:related_work}
Kapoor and Narayanan~\cite{Kapoor:Patterns23} systematically examine
the leakage problem in machine learning-based science. The authors
provide a taxonomy of data leakage with eight categories. The train-test
leakage discussed in this paper belongs to the category \textit{[L1.4] Duplicates in
  datasets}. The authors also conducted a comprehensive
survey of literature and found 17 fields, computer security included,
where researchers have published papers revealing data leakage problems
in machine learning-based science. The security paper cited by Kapoor and Narayanan
is Arp~et.~al~2022~\cite{Arp:USENIXSeurity22}, which covers all the
other seven types of leakage except \textit{[L1.4]}. We have also
extensively searched the computer security literature. The only mentioning
of train-test leakage we could find is from a recent manuscript by
Ding~et.~al~\cite{PrimeVul:arXiv2024}, in the context of vulnerability
datasets for evaluating code language models. The authors identify
both train-test leakage (called ``Code Copy'') and temporal leakage
(called ``Time Travel'') in those datasets. The paper shows the percentage
of train-test leakage in these datasets, but did not conduct experiments
to specifically analyze these leakages' impact on evaluating code language models'
performance. Allamanis~\cite{Allamanis:SIGPLAN19} analyzed the adverse
effects of code duplication in machine learning models of code. While
not explicitly using the term ``data leakage,'' the ``cross-set duplicates''
described in the paper is essentially train-test leakage. There are
also works in other domains as discussed earlier~\cite{Elangovan:arXiv2021, Tampu:SD22, Florensa:NARGAB24}
that explicitly use the term ``data leakage'' to describe this
phenomenon.

Monlina et al.~\cite{Molina:Computers23} pointed out the redundancy
problem in datasets used for machine learning based Android malware
detection. The authors conduct experiments on the impact of duplicate
samples in both training and testing data sets, but did not analyze
the impact of duplicates {\it between} training and testing set,
which is the train-test leakage we focus on. The authors through
experiments point out that duplicates in training data result in
biased model, whereas duplicates in testing data result in biased
evaluation result that disproportionately emphasizes prediction results on
those duplicate samples. The authors used an approach
proposed by Surendran~\cite{Surendran:arxiv21} to filter the dataset
-- removing duplicates and retaining only one representative for each duplicate
group, so that the trained model can generalize better and the
measured performance can better indicate generalization ability of
detectors. Our work examines the data leakage between training and
testing sets, which is an
orthogonal problem to duplicates within training and testing sets.
Our results thus complement those from this prior work.

The train-test leakage discussed in this paper arises from the
presence of identical or nearly identical app representations for distinct apps.
This strongly indicates that the raw APK files share significant
similarities.
Extensive studies have previously delved into Android
app similarities~\cite{Zhou:CODASPY12, Hanna:DIMVA12, Desnos:HICSS12,
Crussell:ESORICS13, Sun:IFIPAICT14, Gonzalez:ICST14, Linares:ICPC16,
Li:ICESS17, Li:JCST19}, with the primary focus of similarity
measurement on localizing and detecting changes resulting from app
repackaging~\cite{Zhou:CODASPY12}, targeting app clones or code
clones~\cite{Crussell:ESORICS13}, identifying changes among app
versions to verify valid code updates~\cite{Desnos:HICSS12}, and
detecting vulnerable code reuse in malware
variants~\cite{Hanna:DIMVA12}.

\section{Conclusion}
\label{sec:conclusion}

This work highlights a prevalent issue of train-test data leakage in
machine learning (ML)-based Android malware detection. We demonstrate
this issue through two case studies, illustrating that such leakage can
inflate model performance and sometimes also lead to misleading research
conclusions. The key issue is that the existence of train-test leakage 
causes the evaluation results to conflate a model's ability to generalize with
its ability to memorize. Thus it is more meaningful to evaluate a machine
learning model's performance on testing data with the leakage removed, than
on the complete testing data. We further use a leak-aware detector scheme
to 
approximate the effect of combining machine learning
with signature-based detection, and shows more clearly which model
can provide more practical utility in real-world use scenarios.

\bibliographystyle{IEEEtran}
\bibliography{android}

\appendix


\vspace{.1in}

\noindent \textbf{Appendix A:} Evaluation Metrics Definition

\vspace{.1in}

\noindent First some common terminologies:

\vspace{.1in}

\noindent \textit{TP}: True Positives (correctly classified malicious apps)

\noindent \textit{FN}: False Negatives (malware mis-classified as benign)

\noindent \textit{TN}: True Negatives (correctly classified benign apps)

\noindent \textit{FP}: False Positives (benign apps mis-classified as malware)

\vspace{.1in}

False Positive Rate (FPR) and False Negative Rate (FNR) quantify the rate at
which the the  model incorrectly classify a sample:

 $$ \textit{FPR} = \frac{\textit{FP}}{\textit{TN} + \textit{FP}} \hspace{.2in} 
 \textit{FNR} = \frac{\textit{FN}}{\textit{TP} + \textit{FN}} $$

FNR and FPR are insensitive to the class ratio between malicious and benign samples.
In practice the small ratio of malware relative to benign apps makes these
metrics unsuitable to reflect a model's real utility, due to the famous
base-rate fallacy~\cite{axelsson:ccs99}.
 For such imbalanced data, precision and recall, and their combination F1 score,
are more suitable.

 \[
   \textit{Precision} = \frac{\textit{TP}}{\textit{TP} + \textit{FP}} \hspace{.2in} 
   \textit{Recall} = \frac{\textit{TP}}{\textit{TP} + \textit{FN}}
 \]

 \[
\textit{F1-score} = 2 \cdot \frac{\textit{Precision} \cdot \textit{Recall}}{\textit{Precision} + \textit{Recall}}
\]

\[ \textit{Sensitivity} = 1 - \textit{FNR} \hspace{.2in} \textit{Specificity} = 1 - \textit{FPR} \]
\[
 \textit{Balanced Accuracy (BA)} = \frac{\textit{Sensitivity} + \textit{Specificity}}{2}
\]

\vspace{.1in}

\noindent \textbf{Appendix B:} Details about Our GCN Model

\vspace{.1in}

 Our GCN model utilizes a four-layer graph convolutional architecture followed by a standard linear layer from the \textit{torch.nn} library. \textbf{ReLU} (Rectified Linear Unit) activation ($max(x, 0)$) was employed in the convolutional layers to extract localized node embeddings.  
 We experimented with varying the number of convolutional layers (2-7) and hidden channels (32-128) within the model. While increasing hidden channels improved efficiency, GPU memory limitations prevented exploration beyond that. We found optimal performance with four convolutional layers and 64 hidden channels.
 Hyperparameter tuning identified 200 epochs and a learning rate of 0.001 as optimal for our dataset.

\vspace{.1in}

\noindent \textbf{Appendix C:} Embedding Similarity Threshold

\vspace{.1in}

In Section~\ref{sec:case_ours}, $M$ defines the level of similarity
between graph embeddings for the purpose of considering train-test
leakage. The following details the process of establishing this
threshold for each data window.

For a given data window, an $M$ value will give rise to a subset of
apps $\mathbf{Leak_{\textit{emb}}}$  in the testing data that are considered
leakage for the GCN model. On the other hand, the subset $\mathbf{Leak_{\textit{fv}}}$
captures the leakage for the RF model. Intuitively, being leakage means that
a testing app is very similar to a training app, whether under the RF model
or the GCN model. Thus we'd like to choose an $M$ so that the corresponding
subset $\mathbf{Leak_{\textit{emb}}}$ most closely matches the subset $\mathbf{Leak_{\textit{fv}}}$.
We use the metric Intersection over Union~(IoU) to measure how closely the two
sets are aligned:

\[
\textit{Intersection over Union} \left( \textit{IoU} \right) = \frac{\mathbf{Leak_{\textit{fv}}} \cap \mathbf{Leak_{\textit{emb}}}}{\mathbf{Leak_{\textit{fv}}} \cup \mathbf{Leak_{\textit{emb}}}}
\]


IoU represents the proportion of elements that are common to $\mathbf{Leak_{\textit{fv}}}$ and $\mathbf{Leak_{\textit{emb}}}$ relative to
their total combined size. It is a normalized score between 0 and 1. 
When the two sets completely coincide, IoU is $1$. When the two sets are disjoint, IoU is $0$.

We follow the steps below to find the threshold $M$ that maximizes IoU:
\begin{enumerate}
\item Define a similarity range (e.g., [0.8, 1.0]).
\item Set a small increment rate (e.g., 0.01) to step through the range.
\item At each step, calculate the IoU based on the current threshold value $M$.
\item We identify the $M$ that leads to the maximum IoU. 
\end{enumerate}

\begin{figure}[h]
  \centering
  \includegraphics[width=0.5\textwidth]{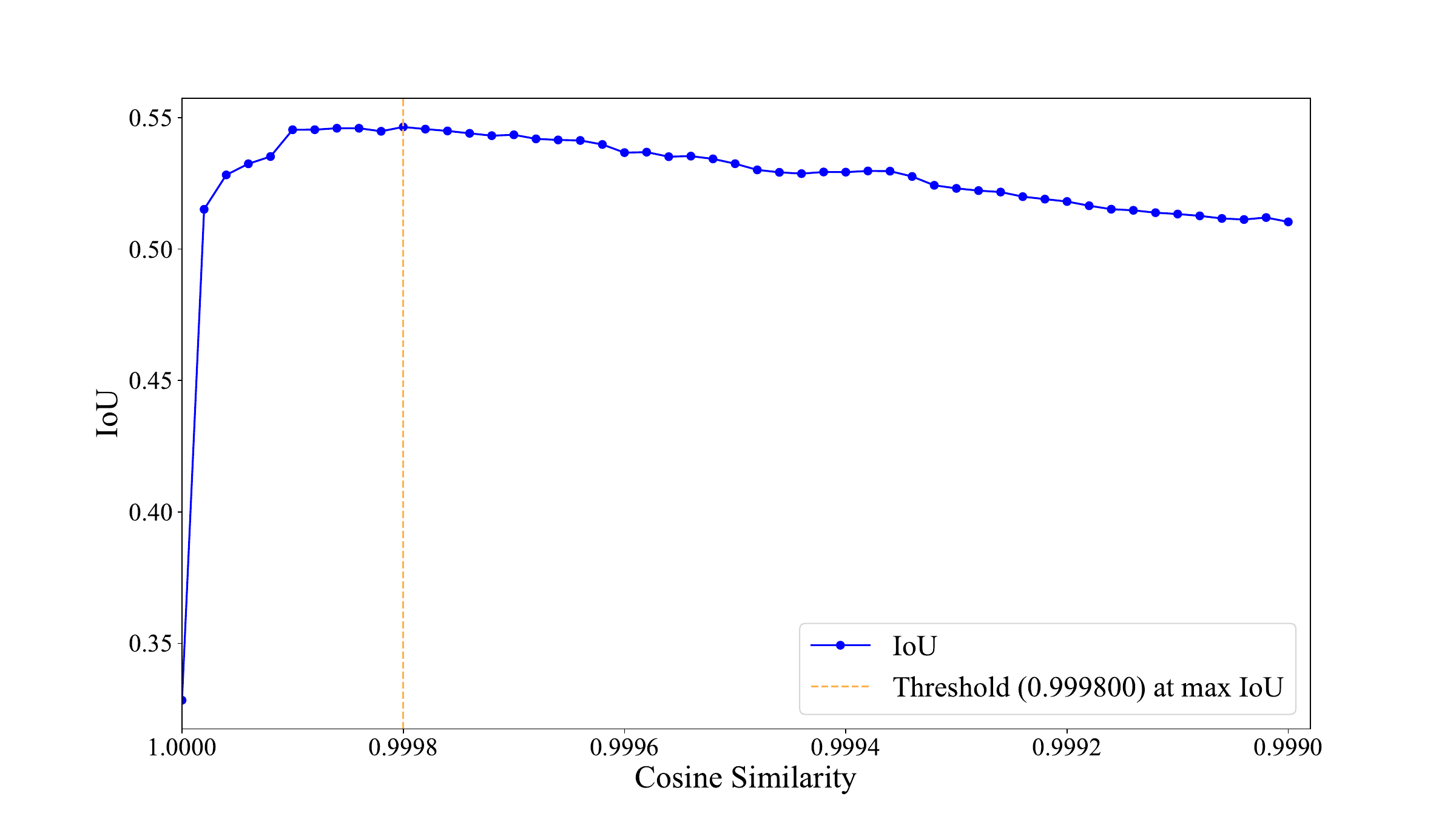}
  \caption{Determining the cosine similarity threshold ($M$) within the experiment window. $M$ is the value that leads to the maximum IoU.}
  \label{fig:cos_threshold}
\end{figure}

The above steps can be achieved through Algorithm~\ref{alg:cos_threshold}. Figure~\ref{fig:cos_threshold} illustrates how we
find thethreshold $M$ for a given experiment window.

\begin{algorithm}[h]
\caption{Finding the Optimal Threshold $M$}\label{alg:cos_threshold}
\begin{algorithmic}[1]
    \State Define a similarity range $R$ (e.g., [0.8, 1.0]).
    \State Set a small increment rate \textit{r} (e.g., 0.01) to step through the range.
    \State Initialize $M_{\text{max}}$ to $0$, $\text{IoU}_{\text{max}}$ to $0$
    \For{$M$ in range from $R_{\text{min}}$ to $R_{\text{max}}$ with step $r$}
        \State Calculate IoU from current $M$ and \textbf{$Leak_{\textit{emb}}$}.
        \If{$\text{IoU} > \text{IoU}_{\text{max}}$}
            \State $\text{IoU}_{\text{max}}$ = $\text{IoU}$
            \State $M_{\text{max}}$ = $M$
        \EndIf
    \EndFor
    \State \textbf{Output:} The optimal threshold $M_{\text{max}}$ that leads to the maximum IoU.
\end{algorithmic}
\end{algorithm}

\begin{table}[h]
  \centering
  \resizebox{\columnwidth}{!}{
  \begin{tabular}{c|c|c|c|c|c|c|c|c}
  \toprule[1.6pt]
    \multicolumn{1}{l|}{} & \textbf{Win\_1} & \textbf{Win\_2} & \textbf{Win\_3} & \textbf{Win\_4} & \textbf{Win\_5} & \textbf{Win\_6} & \textbf{Win\_7} & \textbf{Win\_8} \\
    \hline\hline
    \textbf{Default Test} & 8,000   & 8,000   & 8,000   & 8,000   & 8,000   & 8,000   & 8,000   & 8,000  \\
    $\mathbf{Leak_{\textit{fv}}}$                    & 3,245   & 3,001   & 3,134   & 3,865   & 4,297   & 3,030   & 2,132   & 2,552  \\
    $\mathbf{Leak_{\textit{emb}}}$                    & 4,141   & 3,585   & 4,181   & 4,701   & 5,514   & 3,180   & 3,271   & 4,364  \\
    $\mathbf{Leak_{\textit{fv}} \cup Leak_{\textit{emb}}}$             & 4,694   & 4,203   & 4,929   & 5,362   & 6,066   & 4,529   & 4,395   & 5,256  \\
   \bottomrule[1.6pt]
  \end{tabular}
  }
  \caption{Statistics of test data on each window. $\mathbf{Leak_{\textit{emb}}}$ is determined for each experiment window by selecting the $M$ value that maximizes the IoU score.}
  \label{tab:overlap_info}
\end{table}

\begin{table}[hbt!]
  \centering
  \resizebox{\columnwidth}{!}{
    \begin{tabular}{c|cccccccc}
      \toprule[1.6pt]
      \multirow{2}{*}{\diagbox{\textbf{Train}}{\textbf{Test}}} &  \multirow{2}{*}{\textbf{Win\_1}} &  \multirow{2}{*}{\textbf{Win\_2}} &  \multirow{2}{*}{\textbf{Win\_3}} &  \multirow{2}{*}{\textbf{Win\_4}} &  \multirow{2}{*}{\textbf{Win\_5}} &  \multirow{2}{*}{\textbf{Win\_6}} &  \multirow{2}{*}{\textbf{Win\_7}} &  \multirow{2}{*}{\textbf{Win\_8}} \\
        & & & & & & & & \\
        \hline
        \textbf{Win\_1} & 0.4056 & 0.2699 & 0.1719 & 0.1596 & 0.1535 & 0.1301 & 0.0568 & 0.0321 \\
        \textbf{Win\_2} & & 0.3751 & 0.2256 & 0.2116 & 0.1900 & 0.1585 & 0.0636 & 0.0405 \\
        \textbf{Win\_3} & & & 0.3917 & 0.3275 & 0.2670 & 0.1874 & 0.0811 & 0.0549 \\
        \textbf{Win\_4} & & & & 0.4831 & 0.3698 & 0.2285 & 0.1140 & 0.0620 \\
        \textbf{Win\_5} & & & & & 0.5371 & 0.2850 & 0.1295 & 0.0770 \\
        \textbf{Win\_6} & & & & & & 0.3787 & 0.1565 & 0.1003 \\
        \textbf{Win\_7} & & & & & & & 0.2665 & 0.1224 \\
        \textbf{Win\_8} & & & & & & & & 0.3190 \\
        \bottomrule[1.6pt]
    \end{tabular}
    }
    \caption{\textbf{$Leak_{\textit{fv}}$} ratio in different test window while keep the training data fixed. A decreasing ratio of \textbf{$Leak_{\textit{fv}}$} as the test data release timeline progresses.}
    \label{tab:overlap_changes}
  \end{table}

Table~\ref{tab:overlap_info} summarizes the appearance of $\textit{Leak}_{\textit{fv}}$ or $\textit{Leak}_{\textit{emb}}$ in our test data.
Table~\ref{tab:overlap_changes} shows a consistent decline in the $\textit{Leak}_{\textit{fv}}$ ratio within the test data over time, considering a fixed training dataset. For instance, training models solely on window-1 data leads to progressively lower $\textit{Leak}_{\textit{fv}}$ occurrences in newer test datasets. The $\textit{Leak}_{\textit{emb}}$ ratio follows a similar trend.

\balance

\vspace{.1in}

\noindent \textbf{Appendix D:} Why So Much Leakage?

\vspace{.1in}

Multiple factors could contribute to identical or nearly identical Android app representations. One is the rapid iteration of app
versions. The other reason is the region-based app release: one app
could have multiple language versions for different regions.  While
many APK files are distinct entities based on hash values, they often represent the same application with different versions. Another factor is the reuse of software components during the software development
process. Additionally, the method or data selected to represent the app is another contributing factor.

It is surprising though to see such high percentage of testing apps that share the
same feature vectors with training apps (Figure~\ref{fig:identical_vec_ratio} and \ref{fig:window_identical_vec}),
especially since these models use thousands and even tens of thousands of features.  
To delve deeper into the phenomenon of identical feature vectors, we
manually analyzed four diverse sets of apps. Each set comprised at
least one malicious and one benign app, sharing an identical feature
vector within the set. We randomly selected these sets to ensure they
encompass a representative range of potential scenarios:

\noindent
$S_{1}$ = $\{$ \seqsplit{10f0b79f6e9e10f56e32778e04c62002962039e12fff4ecada7e13fb57c9c025.apk, \\ e9060bf176250385b264777dc49ec5fea6a8df7ce2d52fc103553fca0b5c6f41.apk, \\ \dots}
$\}$

\noindent
$S_{2}$ = $\{$ \seqsplit{1131c3591d6c3788cd3237e2561b1ae3d8fdcb049550d3a57d12f50074a24833.apk, \\ 293652541d80f4b9ecb0b60a9cd89abeff5c7d04d909bcd04cefef7ba6e25478.apk, \\ \dots}
$\}$

\noindent
$S_{3}$ = $\{$ \seqsplit{02f41e035e7ba92896194ee1d3d4274f6226c9fd8e492249ce0e2da505478e4e.apk, \\ 4cfbe089ab61c4254308375e9a3fcf13eaa52f22d7810aecaf09ab7b66a6984d.apk, \\ \dots}
$\}$

\noindent
$S_{4}$ = $\{$ \seqsplit{2c53648d5218a8c76fa9c11d49f9b1764042b6631db03516c30ce90d966103ea.apk, \\ 9215b52ed735b38d4fb91fe3f12af4d2d18f596752a028aeebaf4049e8dca464.apk, \\ \dots}
$\}$

\begin{sloppypar}\noindent For example, in set $S_1$,
\textit{\seqsplit{10f0b79f6e9e10f56e32778e04c62002962039e12fff4ecada7e13fb57c9c025.apk}}
is for paperfolding crafts which is deemed benign by each security
vendor on VirusTotal. \textit{\seqsplit{e9060bf176250385b264777dc49ec5fea6a8df7ce2d52fc103553fca0b5c6f41.apk}}
is for a Barcelona radio station, which is flagged malicious by 9 (out
of 64) security vendors on VirusTotal. However, the two apps have identical
feature vector representations with vector size 2,970.
\end{sloppypar}

The identical feature vectors observed in sets $S_1$, $S_2$, and $S_4$ can be
attributed to a common factor: all the apps in these sets were
developed using the Seattlecloud app development framework
(https://sc.mobileappframeworks.com/). This framework apparently leads
to significant similarities in manifest files and APIs across apps,
even those with diverse functionalities, such as a paperfolding-crafts
app and a Spanish radio station (examples from set 1). Our
investigation revealed that the primary difference between these apps
built using the same platform lies in their HTML code, which is
currently not considered by our feature vector. Notably, some of these
apps utilize inline frames (iframes) within their HTML, which trigger
malicious flags from certain security vendors. Further
investigation is needed to determine whether these flags are false positives.

Following a similar pattern to the previous three sets, all apps in
set $S_3$ are built using the Cordova development framework. Despite
serving distinct purposes (e.g., one app for Arabic scripts/preaching
and another for Perdana 4D, a lottery company), these apps exhibit
remarkably similar manifest files and API calls. Once again, the key
differentiating factor lies in their HTML code. Interestingly, some
apps with suspicious URLs within their HTML trigger ``malicious'' flags
from certain anti-malware vendors. Again, further investigation
is needed to determine if these flags are false positives.

Our analysis reveals a growing trend: developers are increasingly
utilizing well-established app development frameworks to enhance
development efficiency. This trend, however, has an unintended
consequence: apps with diverse functionalities can end up exhibiting
remarkably similar manifest files, API calls, and even bytecode. The
key functionalities of these apps are primarily implemented through
their HTML code, while the bytecode serves more as a common framework
for the HTML code to execute its specific functions within each app.

This section only reports our anectdotal findings on the causes of
the significant train-test leakage in Android app datasets. More
thorough study is needed to fully understand all causes and is
beyond the scope of this paper. Our results do call for awareness
of the train-test leakage phenomenon in this problem domain, and
necessary steps be taken to remove the leakage's impact on performance
evaluation. Moreover, it is perhaps time for a more careful consideration
on how to represent Android apps for machine learning based malware
detection, given that apps with distinct functionalities could end
up with very similar representations under the scheme commonly used
by the current research methods.



\end{document}